\documentclass{aa}
\usepackage{natbib}
\usepackage[colorlinks=true,citecolor=blue]{hyperref}
\usepackage{graphicx}
\usepackage{txfonts}

\def\neworcid#1{\kern .08em\href{https://orcid.org/#1}{\includegraphics[keepaspectratio,width=0.7em]{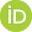}}}
\begin{document}

   \title{Detection of Na in the atmosphere of the hot Jupiter HAT-P-55b}

   \subtitle{}

  \author{
  Huiyi Kang\inst{\ref{pmo},\ref{ustc}}\neworcid{0000-0003-0312-5397} 
  \and Guo Chen\inst{\ref{pmo},\ref{cfe-cp}}\neworcid{0000-0003-0740-5433} 
  \and Chengzi Jiang\inst{\ref{pmo},\ref{ustc}}\neworcid{0000-0003-1381-5527}
  \and Enric Pall\'e\inst{\ref{iac},\ref{ull}}\neworcid{0000-0003-0987-1593} 
  \and Felipe Murgas\inst{\ref{iac},\ref{ull}}\neworcid{0000-0001-9087-1245}
  \and Hannu Parviainen\inst{\ref{ull},\ref{iac}}\neworcid{0000-0001-5519-1391}
  \and Yuehua Ma\inst{\ref{pmo}}
  \and Akihiko Fukui\inst{\ref{komaba},\ref{iac}}\neworcid{0000-0002-4909-5763}
  \and Norio Narita\inst{\ref{komaba},\ref{abc},\ref{iac}}\neworcid{0000-0001-8511-2981}
         }
   \institute{
   CAS Key Laboratory of Planetary Sciences, Purple Mountain Observatory, Chinese Academy of Sciences, No. 10 Yuanhua Road, Qixia District, 210023 Nanjing, PR China\label{pmo} \\
   \email{guochen@pmo.ac.cn}
   \and School of Astronomy and Space Science, University of Science and Technology of China, No. 96 Jinzhai Road, Baohe District, 230026 Hefei, PR China\label{ustc}
   \and CAS Center for Excellence in Comparative Planetology, No. 96 Jinzhai Road, Baohe District, 230026 Hefei, PR China\label{cfe-cp}
   \and Instituto de Astrof\'isica de Canarias (IAC), V\'ia L\'actea s/n, 38205 La Laguna, Tenerife, Spain\label{iac}
   \and Departamento de Astrof\'isica, Universidad de La Laguna (ULL), C/ Padre Herrera, 38206 La Laguna, Tenerife, Spain\label{ull}
   \and Komaba Institute for Science, The University of Tokyo, 3-8-1 Komaba, Meguro, Tokyo 153-8902, Japan\label{komaba}
   \and Astrobiology Center, 2-21-1 Osawa, Mitaka, Tokyo 181-8588, Japan\label{abc}
   }
   \date{Received ...; Accepted ...}

 
  \abstract
  {
  The spectral signatures of optical absorbers, when combined with those of infrared molecules, play a critical role in constraining the cloud properties of exoplanet atmospheres. We aim to use optical transmission spectroscopy to confirm the tentative color signature previously observed by multiband photometry in the atmosphere of hot Jupiter HAT-P-55b. We observed a transit of HAT-P-55b with the OSIRIS spectrograph on the Gran Telescopio Canarias (GTC). We created two sets of spectroscopic light curves using the conventional band-integrated method and the newly proposed pixel-based method to derive the transmission spectrum. We performed Bayesian spectral retrieval analyses on the transmission spectrum to interpret the observed atmospheric properties. The transmission spectra derived from the two methods are consistent, both spectrally resolving the tentative color signature observed by MuSCAT2. The retrievals on the combined OSIRIS and MuSCAT2 transmission spectrum yield the detection of Na at 5.5$\sigma$ and the tentative detection of MgH at 3.4$\sigma$. The current optical-only wavelength coverage cannot constrain the absolute abundances of the atmospheric species. Space-based observations covering the molecular infrared bands or ground-based high-resolution spectroscopy are needed to further constrain the atmospheric properties of HAT-P-55b.
  }


   \maketitle
%

\section{Introduction}
The atmospheres of hot Jupiters can uncover hints of the planet's formation history through atmospheric properties such as metallicity and elemental ratios \citep[e.g.,][]{2011ApJ...743L..16O,2014ApJ...794L..12M,2016ApJ...832...41M,2021ApJ...914...12L}. Among the various observational techniques that could constrain these atmospheric properties, transmission spectroscopy \citep{2000ApJ...537..916S} has been the most widely used, performed on dozens of exoplanets to study the absorption and scattering signatures present in the atmosphere at the day-night terminator. The introduction of retrieval methods \citep[e.g.,][]{2009ApJ...707...24M} can extract the atmospheric properties from observed spectra based on certain assumed parametric forward atmospheric models. 

However, the atmospheric retrieval of transmission spectra could suffer from potential degeneracies \citep{2008A&A...481L..83L,2012ApJ...753..100B,2013Sci...342.1473D,2014RSPTA.37230086G,2017MNRAS.467.2834B,2017MNRAS.470.2972H}, such as the degeneracy between chemical abundances and cloud properties, if the observations were made in a limited spectral range. \citet{2019AJ....157..206W} showed that including optical wavelengths, which are sensitive to clouds and hazes, along with infrared wavelengths could mitigate the degeneracy between chemical abundances and cloud properties. On the other hand, to accurately constrain the properties of clouds and hazes, it is necessary to rely on the detection and characterization of specific spectral features \citep{2023A&A...675A..62J}. 

In particular, Na and K, the dominant optical opacity sources of typical hot Jupiters \citep{2010ApJ...709.1396F}, could exhibit pressure broadened line profiles depending on the cloudiness of the atmosphere, which play a critical role in tightly constraining molecular abundances \citep[e.g.,][]{2019ApJ...887L..20W,2022AJ....164..173C}. Recent low-resolution ground- and space-based transmission spectroscopy surveys \citep[e.g.,][]{2011A&A...527A..73S,2017A&A...600L..11C,2018A&A...616A.145C,2020A&A...642A..54C,2022AJ....164..173C,2018Natur.557..526N,2019AJ....157...21P,2022MNRAS.510.4857A,2023Natur.614..670F} have begun to manifest the line profile of Na and K with tens of nearly uniform narrow bands ($\leq$10~nm), laying the basis for improving future atmospheric retrievals from optical to infrared.

Here we present a follow-up transit observation to obtain the transmission spectrum of the hot Jupiter HAT-P-55b using the Optical System for Imaging and low-Intermediate-Resolution Integrated Spectroscopy \citep[OSIRIS;][]{2000SPIE.4008..623C} mounted on the 10.4~m Gran Telescopio Canarias (GTC). HAT-P-55b is a hot Jupiter orbiting a sun-like star discovered by \citet{2015PASP..127..851J}. Based on four transits observed with the MuSCAT2 four-color simultaneous camera on the 1.52~m Telescopio Carlos S\'{a}nchez (TCS), \citet{2024MNRAS.528.1930K} improved the estimate of the transit and physical parameters for the HAT-P-55 system, obtaining a radius of $1.324^{+0.023}_{-0.022}$~$R_\mathrm{Jup}$, a mass of $0.596^{+0.073}_{-0.072}$~$M_\mathrm{Jup}$, and an equilibrium temperature of $1367\pm 11$~K. The derived MuSCAT2 broadband transmission spectrum in the $g$, $r$, $i$, and $z_s$ was not completely flat, showing moderate evidence that the observed variations could originate from the planetary atmosphere. The goal of this study is to use GTC/OSIRIS to spectrally resolve the optical absorbers responsible for the MuSCAT2 transit depth color variations.

This paper is organized as follows. In Sect.~\ref{sec:data}, we summarize the details of the transit observation and data reduction. In Sect.~\ref{sec:analysis}, we describe the light curve analysis and propose a new pixel-based approach to derive the transmission spectrum in addition to the conventional band-integrated method. In Sect.~\ref{sec:Results}, we present the derived transmission spectrum and perform Bayesian spectral retrieval analysis on the available transmission spectra to understand the atmospheric properties. Finally, we draw conclusions and present our discussion in Sect.~\ref{sec:conclusions}.

\section{Observations and data reduction}
\label{sec:data}
A transit event of the hot Jupiter HAT-P-55b was observed with GTC OSIRIS on May 2, 2020 from 00:23 UT to 05:41 UT. For the transit observation, OSIRIS was configured in the long-slit spectroscopic mode with a $40''$ wide slit and the R1000R grism to cover a wavelength range of 510 to 1000 nm. The OSIRIS CCDs were read in 200 kHz readout mode with $2\times2$ pixel binning. Each binned pixel has a spatial scale of $0.254''$ and an average instrumental dispersion of 0.26~nm. The time series consists of single exposures of 30 s. During the observation the sky was mostly clear and the seeing varied between 0.7$''$ and 1.9$''$. The details of the observation are summarized in Table~\ref{tab:obs_summary}.

To calibrate the flux of HAT-P-55, a reference star (2MASS J17371483+2543127) was simultaneously observed in the same slit. The target and reference stars, which have $r$-band magnitudes of 13.060 and 12.926, respectively \citep{2015AJ....150..101Z}, were placed on the same CCD at a distance of 2.18$'$. For wavelength calibration, the HeAr, Ne, and Xe arc lamps were measured through the 1.00$''$ wide slit using the same R1000R grism. 

\begin{table}
\caption[]{Details of transit observation for HAT-P-55b.}
\renewcommand\arraystretch{1.5} 
\begin{center}
\setlength{\tabcolsep}{1.5mm}
\begin{tabular}{lc}
\hline\hline
Target star                 & HAT-P-55\\
Reference star              & 2MASS J17371483+2543127\\
\hline
Date (UT)                   & 2020-05-02\\
Observing time (UT)         & 00:23--05:41\\
Slit ($''$)                 & 40\\
Exposure time (s)           & 30\\
Number of exposures         & 340 \\
Airmass\tablefootmark{a}       & 1.538--1.001--1.070\\
Seeing\tablefootmark{b} ($''$) & 0.7--1.9\\
Median S/N per pixel\tablefootmark{c} & 238--240--242 (Target)\\
  & 252--257--258 (Reference)\\
\hline
\end{tabular}
\tablefoot{
The observation was performed with Proposal ID GTC24-20A (PI: Enric Pall\'e).
\tablefoottext{a}{Start/minimum/end.}
\tablefoottext{b}{95\% interval, measured by the full width at half maximum of the stellar spectrum along the spatial direction at the central wavelength.}
\tablefoottext{c}{Pre-transit/in-transit/post-transit.}
}
\label{tab:obs_summary}  
\end{center}
\end{table}

\begin{figure}
\centering
\includegraphics[width=0.5\textwidth]{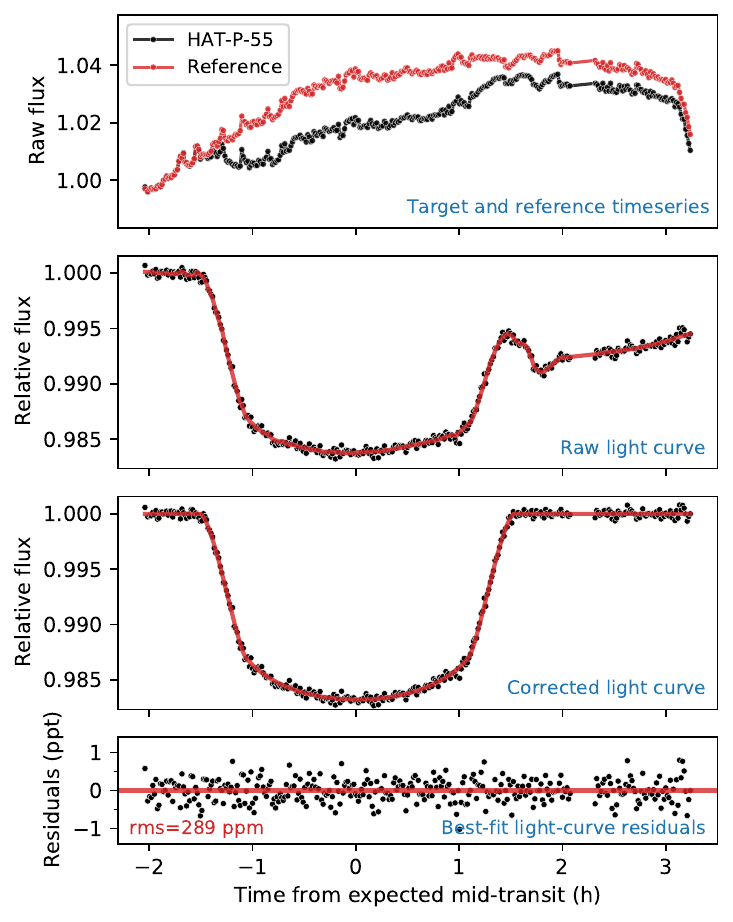}
\caption{Band-integrated white light curve of HAT-P-55 observed with GTC/OSIRIS. The first row shows the flux time series of HAT-P-55 and its reference star. The second row shows the raw light curve after dividing the flux of HAT-P-55 by that of the reference. The red line represents the best-fit combined transit and systematics  model. The third row shows the light curve after removing the systematics. The fourth row shows the best-fit light curve residuals. }
\label{fig: wlc}
\end{figure}

The raw OSIRIS spectral images were reduced using the approaches developed in our previous studies \citep[e.g.,][]{2017A&A...600A.138C,2018A&A...616A.145C,2020A&A...642A..54C,2022A&A...664A..50J}, including calibration of overscan, bias, flat field, cosmic rays, and sky background. In particular, the sky background model was constructed in the wavelength domain based on the two-dimensional pixel-to-wavelength solution derived from the arc lamps. The one-dimensional spectra of the target and reference stars were extracted using the optimal extraction algorithm \citep{1986PASP...98..609H} with an aperture diameter of 42~pixels, optimized to minimize subsequent light curve scatter, and aligned to the laborotary rest frame. The center of the exposure time in UT was calculated for each spectrum and converted to barycentric Julian dates in barycentric dynamical time \citep[$\mathrm{BJD}_\mathrm{TDB}$;][]{2010PASP..122..935E}, which were recorded as time stamps. 

Two methods were used to create the light curves. Since the pixel-to-wavelength transformation solution is nonlinear, the flux within a given passband can be obtained either in the original pixel grid or in the linearly resampled wavelength grid. In our conventional method \citep{2017A&A...600A.138C}, to eliminate the interpolation process, the pixel range for a given passband was calculated and only the flux of the partial pixels at the edge was interpolated, while those of the full pixels were summed directly. Two types of light curves were created. For the band-integrated white light curve, the flux was summed within the wavelength range of 511.5--901.5~nm, excluding 754--768~nm due to the steep absorption feature of the telluric oxygen A-band. The wavelengths longer than 901.5~nm were not used to avoid potential second-order contamination and fringing effects. For the band-integrated spectroscopic light curves, the entire wavelength range was divided into 55 passbands of 5~nm, 1 passband of 15~nm, and 3 passband of 30~nm. Each light curve was normalized by its own out-of-transit flux. 

In addition to the conventional method, we introduced a new method for creating the spectroscopic light curves. The band-integrated white light curve in our new method was the same as in the conventional method. However, the spectroscopic light curves were created in a different way. The flux of the original pixels in each spectrum does not form a valid time series due to misalignment along the time axis and nonlinear wavelength solutions. Therefore, all spectra were resampled using the misalignment corrected wavelength solutions in a linear wavelength grid from 511.5 to 901.5 nm with an average dispersion of 0.26 nm per pixel, taking into account flux conservation. This resulted in 1500 uniform pseudo pixels and thus 1500 pixel light curves. The transit depths derived from the spectroscopic light curves using these two methods are compared in the subsequent analysis.

\section{Light-curve analysis}
\label{sec:analysis}
\subsection{White light curve}
The white light curve was modeled with a transit model configured with the nonlinear limb darkening law and a circular orbit \citep{2015PASP..127..851J} using the Python package {\tt batman} \citep{2015PASP..127.1161K}. The transit model was parameterized by orbital period $P$, radius ratio $R_\mathrm{p}/R_\star$, orbital inclination $i$, scaled semimajor axis $a/R_{\star}$, mid-transit time $T_\mathrm{mid}$, and four limb darkening coefficients (LDCs) $u_i$. The values of $P$, $i$, and $a/R_{\star}$ were fixed to 3.58523130~d, 86.80$^\circ$, and 9.06, respectively, derived in \citet{2024MNRAS.528.1930K}. The four LDCs were fixed to the pre-calculated values ($u_1=1.2247$, $u_2=-2.0154$, $u_3=2.6623$, $u_4=-1.0898$) from the PHOENIX stellar atmosphere models using the Python code of \citet{2015MNRAS.450.1879E}, with stellar effective temperature $T_\mathrm{eff}=5800~{\rm{K}}$, surface gravity $\log g_\star=4.5$, and metallicity $\mathrm{[Fe/H]}=0.0$. 

The correlated systematics in the light curve were described by Gaussian processes \citep[GP; e.g.,][]{2012MNRAS.419.2683G}, assuming the transit model as the mean function. The Python package {\tt george} \citep{2015ITPAM..38..252A} was used to perform the GP regression with the capability of multidimensional inputs in the covariance matrix, which adopted the Matern-3/2 kernel $k_\mathrm{M32}=(1+\sqrt{3r^2})\exp(-\sqrt{3r^2})$ for two points in each input at a given radius $r$. The kernel was a multiplication of the kernels of each individual input, and the inputs to the combined kernel were time $t$, spectral drift $x$, spatial drift $y$, full width at half maximum of the target point spread function $s$, and rotation angle $\theta$. The combined kernel had a scale length hyperparameter $\rho_i$ for each input and a single covariance hyperparameter $\sigma_r^2$. An additional white noise jitter $\sigma_w^2$ was used to account for potential underestimation of white noise. 

To investigate the posterior probability distributions of the free parameters, the affine invariant Markov Chain Monte Carlo (MCMC) ensemble sampler was implemented using the Python package {\tt emcee} \citep{2013PASP..125..306F}. Uniform priors were used for $T_\mathrm{mid}$ and log-uniform priors for the GP hyperparameters. Given the presence of strong correlated noise distorting the white light curve, a normal prior of $0.1220\pm0.0007$ from \citet{2024MNRAS.528.1930K} was imposed on $R_\mathrm{p}/R_\star$ to better constrain the correlated systematics, useful for extracting the common mode trend. For the MCMC process, two short chains were run for the burn-in phase and one long chain to ensure convergence for the posterior estimation. 

Table~\ref{tab: transit_param} gives a summary of the assumed priors and estimated posteriors for the parameters used in the modeling of the white light curve. Figure~\ref{fig: wlc} shows the white light curve and the best-fit model. The best-fit light curve residuals have a standard derivation of 289~ppm, which is 1.94 times the expected photon noise.

\begin{table}
\renewcommand\arraystretch{1.5}
\centering
\caption{Parameters derived from the white light curve.}
\label{tab: transit_param}
\scalebox{1.}{
\begin{tabular}{ccc}
\hline\hline
Parameter	& Prior & Posterior\\
\hline
$P$ (d) & 3.58523\tablefootmark{a}  & -- \\
$i$ ($^{\circ}$) & 86.80\tablefootmark{a} & --\\
$a/R_{\star}$ & 9.06\tablefootmark{a}  & --\\
$T_\mathrm{mid}$ (MJD\tablefootmark{b}) & $\mathcal{U}(0.59,0.63)$ & $0.60580^{+0.00019}_{-0.00018}$\\
$R_p/R_{\star}$ & $\mathcal{N}(0.1220,0.0007)$ & $0.12199^{+0.00068}_{-0.00068}$\\
$u_1$ & 1.2247 & --\\
$u_2$ & $-$2.0154 & --\\
$u_3$ & 2.6623 & --\\
$u_4$ & $-$1.0898 & --\\
$\ln \sigma_w$ & $\mathcal{U}(-16,-5)$ & $-8.246^{+0.053}_{-0.054}$\\
$\ln \sigma_r$ & $\mathcal{U}(-10,-1)$ & $-6.21^{+0.36}_{-0.31}$\\
$\ln \rho_t$ & $\mathcal{U}(-6,5)$ & $-1.55^{+0.41}_{-0.39}$\\
$\ln \rho_x$ & $\mathcal{U}(-5,5)$ & $4.46^{+0.39}_{-0.56}$\\
$\ln \rho_y$ & $\mathcal{U}(-5,5)$ & $4.22^{+0.55}_{-0.79}$\\
$\ln \rho_s$ & $\mathcal{U}(-5,5)$ & $3.69^{+0.61}_{-0.49}$\\
$\ln \rho_\theta$ & $\mathcal{U}(-5,5)$ & $0.54^{+0.31}_{-0.29}$\\
\hline
\end{tabular}
}
\tablefoot{
\tablefoottext{a}{Taken from \citet{2024MNRAS.528.1930K}.}
\tablefoottext{b}{$\rm{MJD=BJD_{TDB} - 2458971}.$}
}
\end{table}

\begin{figure}
\centering
\includegraphics[width=0.5\textwidth]{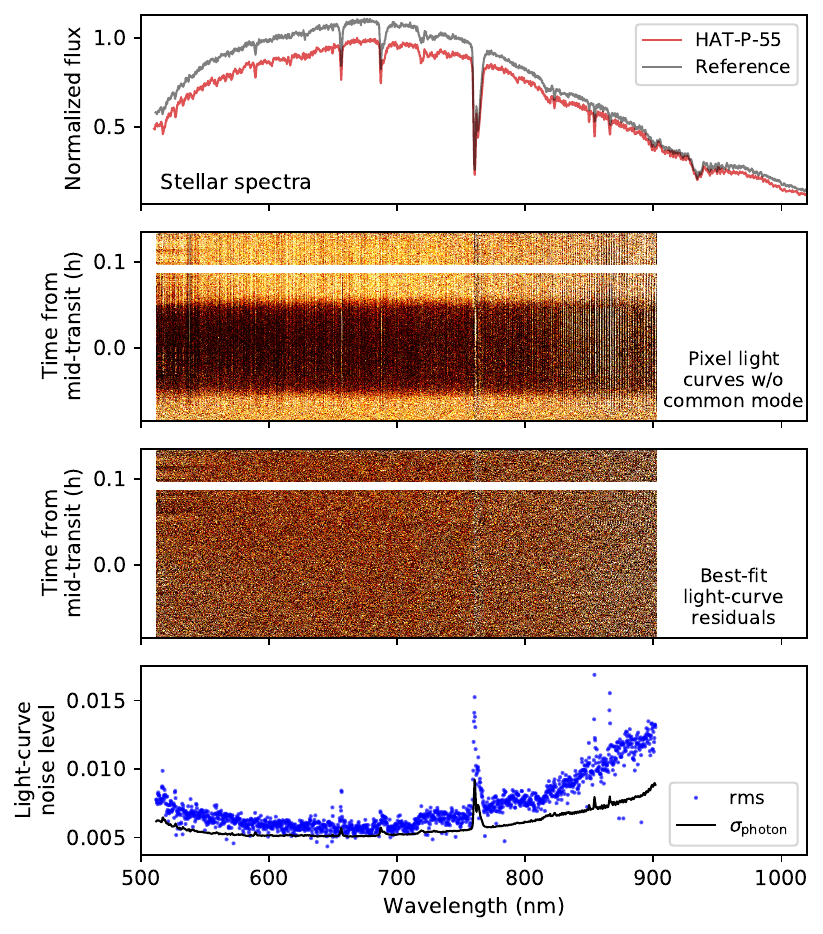}
\caption{Pixel-by-pixel spectroscopic light curves of HAT-P-55 observed with GTC/OSIRIS. The first row shows the spectra of HAT-P-55 and the reference star. The second row shows the matrix of pixel light curves after removing the common mode systematics. The third row shows the matrix of the best-fit light curve residuals. The fourth row shows the rms of each pixel light curve, compared to the expected photon noise.}
\label{fig: pixel_light_curves}
\end{figure}

\begin{table*}
\renewcommand\arraystretch{1.2}
\caption[]{Transmission spectra of HAT-P-55b obtained by the conventional method and the new method.}
\begin{center}
\def\tol#1#2#3{\hbox{\rule{0pt}{15pt}${#1}^{{#2}}_{{#3}}$}}
\setlength{\tabcolsep}{1.5mm}
\scalebox{0.85}{
\begin{tabular}{ccccccc}
\hline\hline
$\lambda$ (nm) & $u_1$ & $u_2$ & $u_3$ & $u_4$ & $R_{\rm{p}}/R_\star|_\mathrm{conventional}$ &  $R_{\rm{p}}/R_\star|_\mathrm{new}$ (adopted)\\\hline
    511.5--516.5 & $0.9599$ & $-1.5402$ & $2.5456$ & $-1.0873$ & $0.12528 ^{+0.03114}_{-0.03382}$ & $0.12564 ^{+0.00165}_{-0.00165}$\\
    516.5--521.5 & $0.9351$ & $-1.3971$ & $2.2707$ & $-0.9598$ & $0.12350 ^{+0.00246}_{-0.00281}$ & $0.11952 ^{+0.00197}_{-0.00197}$\\
    521.5--526.5 & $0.9261$ & $-1.3266$ & $2.2291$ & $-0.9642$ & $0.12846 ^{+0.00389}_{-0.00541}$ & $0.12596 ^{+0.00147}_{-0.00147}$\\
    526.5--531.5 & $0.9860$ & $-1.4900$ & $2.3611$ & $-0.9972$ & $0.12590 ^{+0.00229}_{-0.00484}$ & $0.12501 ^{+0.00226}_{-0.00226}$\\
    531.5--536.5 & $0.9739$ & $-1.4682$ & $2.3587$ & $-1.0032$ & $0.12415 ^{+0.00138}_{-0.00192}$ & $0.12173 ^{+0.00125}_{-0.00125}$\\
    536.5--541.5 & $0.9647$ & $-1.4313$ & $2.2638$ & $-0.9544$ & $0.12646 ^{+0.00410}_{-0.00616}$ & $0.12176 ^{+0.00229}_{-0.00229}$\\
    541.5--546.5 & $0.9461$ & $-1.3672$ & $2.2322$ & $-0.9614$ & $0.12519 ^{+0.00133}_{-0.00157}$ & $0.12473 ^{+0.00121}_{-0.00121}$\\
    546.5--551.5 & $0.9967$ & $-1.5067$ & $2.3386$ & $-0.9812$ & $0.12071 ^{+0.00159}_{-0.00187}$ & $0.12187 ^{+0.00090}_{-0.00090}$\\
    551.5--556.5 & $1.0018$ & $-1.5091$ & $2.3490$ & $-0.9941$ & $0.12460 ^{+0.00087}_{-0.00147}$ & $0.12408 ^{+0.00130}_{-0.00130}$\\
    556.5--561.5 & $1.0106$ & $-1.5193$ & $2.3656$ & $-1.0084$ & $0.12453 ^{+0.00147}_{-0.00176}$ & $0.12231 ^{+0.00143}_{-0.00143}$\\
    561.5--566.5 & $1.0242$ & $-1.5465$ & $2.3796$ & $-1.0104$ & $0.12440 ^{+0.00103}_{-0.00130}$ & $0.12423 ^{+0.00095}_{-0.00095}$\\
    566.5--571.5 & $1.0212$ & $-1.5200$ & $2.3325$ & $-0.9923$ & $0.12512 ^{+0.00136}_{-0.00118}$ & $0.12299 ^{+0.00178}_{-0.00178}$\\
    571.5--576.5 & $1.0339$ & $-1.5698$ & $2.4024$ & $-1.0217$ & $0.12278 ^{+0.00141}_{-0.00185}$ & $0.12568 ^{+0.00110}_{-0.00110}$\\
    576.5--581.5 & $1.0392$ & $-1.5621$ & $2.3760$ & $-1.0114$ & $0.12560 ^{+0.00103}_{-0.00112}$ & $0.12332 ^{+0.00110}_{-0.00110}$\\
    581.5--586.5 & $1.0429$ & $-1.5441$ & $2.3178$ & $-0.9845$ & $0.12320 ^{+0.00203}_{-0.00215}$ & $0.12483 ^{+0.00086}_{-0.00086}$\\
    586.5--591.5 & $1.0712$ & $-1.6545$ & $2.4480$ & $-1.0307$ & $0.12345 ^{+0.00383}_{-0.00432}$ & $0.12617 ^{+0.00172}_{-0.00172}$\\
    591.5--596.5 & $1.0549$ & $-1.5910$ & $2.3644$ & $-0.9986$ & $0.12516 ^{+0.00091}_{-0.00101}$ & $0.12440 ^{+0.00078}_{-0.00078}$\\
    596.5--601.5 & $1.0711$ & $-1.6200$ & $2.3818$ & $-1.0054$ & $0.12422 ^{+0.00127}_{-0.00145}$ & $0.12373 ^{+0.00095}_{-0.00095}$\\
    601.5--606.5 & $1.0949$ & $-1.6936$ & $2.4609$ & $-1.0336$ & $0.12458 ^{+0.00113}_{-0.00120}$ & $0.12401 ^{+0.00084}_{-0.00084}$\\
    606.5--611.5 & $1.1145$ & $-1.7433$ & $2.4760$ & $-1.0266$ & $0.12365 ^{+0.00102}_{-0.00117}$ & $0.12371 ^{+0.00129}_{-0.00129}$\\
    611.5--616.5 & $1.1414$ & $-1.7973$ & $2.5163$ & $-1.0415$ & $0.12415 ^{+0.00282}_{-0.00233}$ & $0.12138 ^{+0.00110}_{-0.00110}$\\
    616.5--621.5 & $1.1135$ & $-1.7284$ & $2.4594$ & $-1.0262$ & $0.12470 ^{+0.00138}_{-0.00126}$ & $0.12414 ^{+0.00089}_{-0.00089}$\\
    621.5--626.5 & $1.1324$ & $-1.7741$ & $2.4744$ & $-1.0238$ & $0.12305 ^{+0.00214}_{-0.00160}$ & $0.12206 ^{+0.00133}_{-0.00133}$\\
    626.5--631.5 & $1.1429$ & $-1.8096$ & $2.5405$ & $-1.0571$ & $0.12424 ^{+0.00081}_{-0.00083}$ & $0.12226 ^{+0.00161}_{-0.00161}$\\
    631.5--636.5 & $1.1252$ & $-1.7400$ & $2.4388$ & $-1.0151$ & $0.12195 ^{+0.00306}_{-0.00268}$ & $0.12259 ^{+0.00123}_{-0.00123}$\\
    636.5--641.5 & $1.1470$ & $-1.8091$ & $2.5104$ & $-1.0402$ & $0.12407 ^{+0.00085}_{-0.00066}$ & $0.12261 ^{+0.00117}_{-0.00117}$\\
    641.5--646.5 & $1.1465$ & $-1.8112$ & $2.5109$ & $-1.0406$ & $0.12117 ^{+0.00059}_{-0.00056}$ & $0.12056 ^{+0.00114}_{-0.00114}$\\
    646.5--651.5 & $1.1514$ & $-1.8035$ & $2.4834$ & $-1.0294$ & $0.12365 ^{+0.00248}_{-0.00197}$ & $0.12018 ^{+0.00191}_{-0.00191}$\\
    651.5--656.5 & $1.1822$ & $-1.7797$ & $2.3728$ & $-0.9880$ & $0.12041 ^{+0.00338}_{-0.00344}$ & $0.12069 ^{+0.00296}_{-0.00296}$\\
    656.5--661.5 & $1.2055$ & $-1.8702$ & $2.4858$ & $-1.0319$ & $0.12125 ^{+0.00096}_{-0.00083}$ & $0.12127 ^{+0.00170}_{-0.00170}$\\
    661.5--666.5 & $1.1875$ & $-1.8980$ & $2.5709$ & $-1.0622$ & $0.12285 ^{+0.00099}_{-0.00098}$ & $0.12146 ^{+0.00105}_{-0.00105}$\\
    666.5--671.5 & $1.1880$ & $-1.8933$ & $2.5436$ & $-1.0466$ & $0.12052 ^{+0.00402}_{-0.00361}$ & $0.12014 ^{+0.00111}_{-0.00111}$\\
    671.5--676.5 & $1.1917$ & $-1.9017$ & $2.5470$ & $-1.0471$ & $0.12137 ^{+0.00128}_{-0.00124}$ & $0.12032 ^{+0.00063}_{-0.00063}$\\
    676.5--681.5 & $1.1981$ & $-1.9320$ & $2.5812$ & $-1.0588$ & $0.12109 ^{+0.00218}_{-0.00229}$ & $0.11865 ^{+0.00078}_{-0.00078}$\\
    681.5--686.5 & $1.2295$ & $-2.0112$ & $2.6519$ & $-1.0827$ & $0.12404 ^{+0.00323}_{-0.00307}$ & $0.12242 ^{+0.00179}_{-0.00179}$\\
    686.5--691.5 & $1.2213$ & $-1.9883$ & $2.6339$ & $-1.0800$ & $0.11942 ^{+0.00141}_{-0.00188}$ & $0.11929 ^{+0.00305}_{-0.00305}$\\
    691.5--696.5 & $1.2317$ & $-2.0149$ & $2.6550$ & $-1.0868$ & $0.11770 ^{+0.00373}_{-0.00310}$ & $0.11826 ^{+0.00058}_{-0.00058}$\\
    696.5--701.5 & $1.2287$ & $-1.9979$ & $2.6305$ & $-1.0781$ & $0.12120 ^{+0.00191}_{-0.00157}$ & $0.12189 ^{+0.00136}_{-0.00136}$\\
    701.5--706.5 & $1.2476$ & $-2.0505$ & $2.6728$ & $-1.0902$ & $0.12125 ^{+0.00214}_{-0.00198}$ & $0.12107 ^{+0.00101}_{-0.00101}$\\
    706.5--711.5 & $1.2492$ & $-2.0615$ & $2.6849$ & $-1.0947$ & $0.12274 ^{+0.00128}_{-0.00127}$ & $0.12145 ^{+0.00108}_{-0.00108}$\\
    711.5--716.5 & $1.2562$ & $-2.0706$ & $2.6839$ & $-1.0937$ & $0.12358 ^{+0.00237}_{-0.00188}$ & $0.12165 ^{+0.00122}_{-0.00122}$\\
    716.5--721.5 & $1.2697$ & $-2.1126$ & $2.7045$ & $-1.0923$ & $0.12033 ^{+0.00206}_{-0.00329}$ & $0.11931 ^{+0.00101}_{-0.00101}$\\
    721.5--726.5 & $1.2576$ & $-2.0734$ & $2.6833$ & $-1.0949$ & $0.12343 ^{+0.00102}_{-0.00119}$ & $0.12172 ^{+0.00113}_{-0.00113}$\\
    726.5--731.5 & $1.2835$ & $-2.1558$ & $2.7369$ & $-1.1015$ & $0.12114 ^{+0.00212}_{-0.00187}$ & $0.11983 ^{+0.00209}_{-0.00209}$\\
    731.5--736.5 & $1.2724$ & $-2.1173$ & $2.7127$ & $-1.1015$ & $0.12112 ^{+0.00222}_{-0.00113}$ & $0.11963 ^{+0.00096}_{-0.00096}$\\
    736.5--741.5 & $1.2986$ & $-2.1912$ & $2.7868$ & $-1.1280$ & $0.11835 ^{+0.00172}_{-0.00136}$ & $0.11716 ^{+0.00198}_{-0.00198}$\\
    741.5--746.5 & $1.3170$ & $-2.2488$ & $2.8324$ & $-1.1390$ & $0.11965 ^{+0.00303}_{-0.00211}$ & $0.11980 ^{+0.00139}_{-0.00139}$\\
    746.5--751.5 & $1.3070$ & $-2.2144$ & $2.7911$ & $-1.1262$ & $0.12059 ^{+0.00193}_{-0.00237}$ & $0.12084 ^{+0.00154}_{-0.00154}$\\
    751.5--756.5 & $1.3036$ & $-2.2021$ & $2.7837$ & $-1.1257$ & $0.12177 ^{+0.00073}_{-0.00073}$ & $0.11974 ^{+0.00124}_{-0.00124}$\\
    756.5--761.5 & $1.3160$ & $-2.2366$ & $2.8024$ & $-1.1282$ & $0.11746 ^{+0.00477}_{-0.00405}$ & $0.12173 ^{+0.00511}_{-0.00511}$\\
    761.5--766.5 & $1.3082$ & $-2.2137$ & $2.7693$ & $-1.1141$ & $0.12026 ^{+0.00148}_{-0.00168}$ & $0.11287 ^{+0.00329}_{-0.00329}$\\
    766.5--771.5 & $1.3057$ & $-2.1992$ & $2.7540$ & $-1.1113$ & $0.12007 ^{+0.00076}_{-0.00077}$ & $0.12072 ^{+0.00206}_{-0.00206}$\\
    771.5--776.5 & $1.3126$ & $-2.2174$ & $2.7775$ & $-1.1219$ & $0.11983 ^{+0.00268}_{-0.00222}$ & $0.11745 ^{+0.00207}_{-0.00207}$\\
    776.5--781.5 & $1.3418$ & $-2.3051$ & $2.8655$ & $-1.1526$ & $0.12162 ^{+0.00235}_{-0.00191}$ & $0.12164 ^{+0.00121}_{-0.00121}$\\
    781.5--786.5 & $1.3234$ & $-2.2525$ & $2.8120$ & $-1.1349$ & $0.12263 ^{+0.00272}_{-0.00214}$ & $0.12006 ^{+0.00183}_{-0.00183}$\\
    786.5--801.5 & $1.3451$ & $-2.3179$ & $2.8717$ & $-1.1545$ & $0.12093 ^{+0.00164}_{-0.00138}$ & $0.11870 ^{+0.00107}_{-0.00107}$\\
    801.5--831.5 & $1.3643$ & $-2.3663$ & $2.8941$ & $-1.1593$ & $0.12171 ^{+0.00175}_{-0.00183}$ & $0.12034 ^{+0.00159}_{-0.00159}$\\
    831.5--861.5 & $1.4090$ & $-2.4848$ & $2.9903$ & $-1.1924$ & $0.12180 ^{+0.00078}_{-0.00089}$ & $0.11994 ^{+0.00276}_{-0.00276}$\\
    861.5--891.5 & $1.3944$ & $-2.4534$ & $2.9451$ & $-1.1762$ & $0.12192 ^{+0.00112}_{-0.00113}$ & $0.12314 ^{+0.00256}_{-0.00256}$\\
\hline
\end{tabular}
}
\label{tab:transpec}  
\end{center}
\end{table*}

\subsection{Spectroscopic light curves}
Two sets of spectroscopic light curves were produced using the conventional method and the new methods described in Sect.~\ref{sec:data}. Similar to the correlated systematics in the white light curve, there was a common mode trend in the spectroscopic light curves. An empirical model for the common mode trend was derived by dividing the raw white light curve by its best-fit transit model. For the products of both methods, each individual spectroscopic light curve was divided by this empirical model to correct for the common mode systematics before further analysis. 

For the conventional method, the 59 band-integrated spectroscopic curves were modeled in the same way as the white light curve, with the same Matern-3/2 kernel and the same GP inputs, except that the mid-transit time was fixed to the one derived in the white light curve (see Table~\ref{tab: transit_param}) and the radius ratios were fitted with uniform priors $\mathcal{U}(0.07,0.17)$.  

For the new method, the 1500 pixel light curves were also modeled with freely fitted radius ratios with uniform priors and the same fixed mid-transit time. However, unlike the band-integrated spectroscopic light curves, the pixel light curves were dominated by white noise instead of correlated noise. To avoid poor constraints on the GP hyperparameters in the case of high-dimensional GP inputs, only the time vector was used in the covariance kernel of the GP regression of the pixel light curves. In addition, to reduce computational cost, the Python package {\tt celerite} \citep{2017AJ....154..220F} was used to perform the GP regression with an approximated Matern-3/2 kernel. To improve the signal-to-noise ratio, the transit depths of the pixel light curves were binned into the 59 passbands used in the conventional method. 

Table~\ref{tab:transpec} presents the chromatic radius ratios, i.e. the transmission spectra, derived from the spectroscopic light curves produced by the conventional method and the new method, along with the adopted nonlinear limb darkening coefficients. Figure~\ref{fig: pixel_light_curves} shows the matrix of the pixel-by-pixel spectroscopic light curves using the new method along with the best-fit light curve residuals. In general, a light curve precision of 0.76--2.12 times the expected photon noise is achieved when modeling pixel light curves, which is in broad agreement with the range of 1.33--2.07 obtained in conventional band-integrated spectroscopic light curves.

\begin{figure}
\centering
\includegraphics[width=1.0\linewidth]{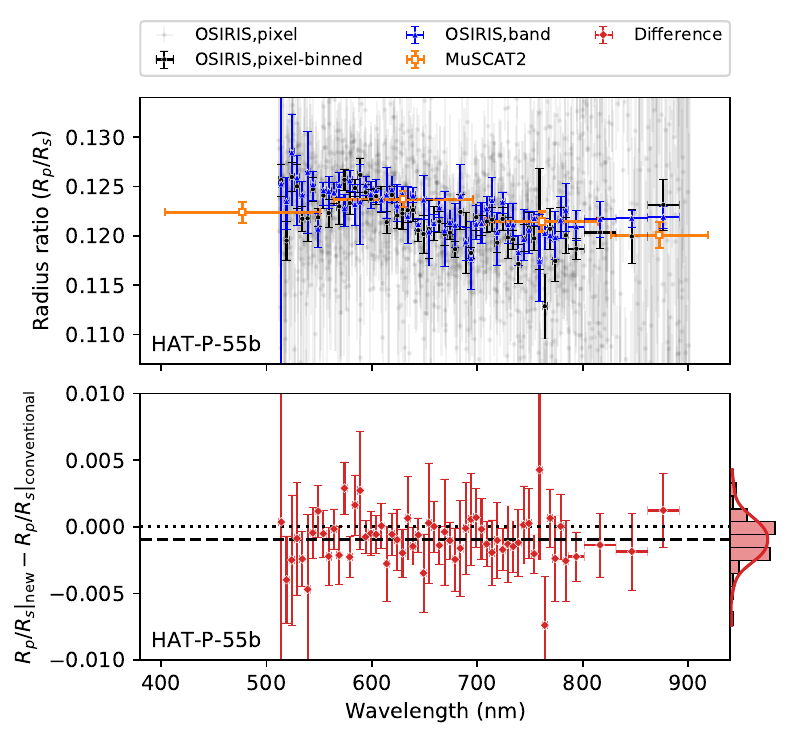}
\caption{The upper panel presents the transmission spectra of HAT-P-55b. The blue triangles and black circles show the OSIRIS transmission spectra obtained with the conventional (denoted by ``band'') and new (denoted by ``pixel-binned'') methods. The gray dots show the pixel transmission spectrum. The orange squares show the broadband transmission spectrum obtained with MuSCAT2 by \citet{2024MNRAS.528.1930K}. The bottom panel presents the difference between the new and conventional methods (red diamonds) and the corresponding histogram. The dashed line marks an overall offset of $\Delta R_\mathrm{p}/R_\star=-0.00094$.}
\label{fig: transpec_comparison}
\end{figure}

\section{Results and interpretations}
\label{sec:Results}
\subsection{Transmission spectrum}
As shown in Fig.~\ref{fig: transpec_comparison}, the GTC/OSIRIS transmission spectra derived from the conventional and new methods have a difference with a chi-squared value of $\chi^2=34.2$ for 59 degrees of freedom (dof), indicating good agreement between the two methods. Allowing a free offset ($\Delta R_\mathrm{p}/R_\star=0.00094\pm 0.00027$) between them would reduce the difference to a chi-squared value of $\chi^2=22.6$ for 58 dof. The largest deviation is measured at 764~nm, which is heavily contaminated by the telluric oxygen A-band. The main difference between the two methods is that the conventional method produces flux-weighted radius ratios and systematics in each spectroscopic band, while the new method derives radius ratios on a pixel basis and then calculates the average values of all pixels in the corresponding band without weighting. However, both methods have resulted in consistent transmission spectra in this work.

In the following analysis, we will focus on the transmission spectrum derived from the new method, excluding the telluric-contaminated bands at 759~nm and 764~nm. This OSIRIS spectrum is significantly not flat at a confidence level of 8.7$\sigma$ ($\chi^2=201.8$ for 56 dof). For comparison, the broadband transmission spectrum obtained by \citet{2024MNRAS.528.1930K} using TCS/MuSCAT2 is also shown in Fig~\ref{fig: transpec_comparison}. Both the OSIRIS and MuSCAT2 transmission spectra show a similar spectral shape, i.e., the radius ratio increases toward 589~nm but decreases for wavelengths longer than 589~nm. This implies that the potential atmospheric spectral signature observed in the MuSCAT2 broadband ``spectrum'' could be due to the Na absorption, as suggested by the resolved spectral shape of the OSIRIS spectrum.

\subsection{Atmospheric retrievals}
\label{sec:retrievals}
\subsubsection{Model setup}
To interpret the combined OSIRIS and MuSCAT2 transmission spectra of HAT-P-55b, we performed spectral retrieval analyses on the observed spectra. We used the Python package {\tt petitRADTRANS} \citep{2019A&A...627A..67M} to create forward models for the planetary atmosphere in the transmission geometry, and the Python package {\tt PyMultiNest} \citep{2014A&A...564A.125B} to estimate model evidence and parameter posterior distributions in the Bayesian framework using the nested sampling algorithm. 

In our models, we assumed an isothermal atmosphere with a temperature of $T_\mathrm{iso}$ and divided the pressure range between $10^2$ and $10^{-6}$~bar into 100 equally spaced layers in logarithmic space. We set the reference pressure $P_0$ as a free parameter, corresponding to the white-light planetary radius ($R_\mathrm{p}=1.324$~$R_\mathrm{J}$). We used the planetary gravity $\log g_\mathrm{p}=2.926$ and the stellar radius $R_\star=1.105$~$R_\sun$ to compute the transmission spectra in terms of planet-to-star radius ratio. The values of the planetary and stellar parameters were taken from \citet{2024MNRAS.528.1930K}. Our models included line opacity sources from gas species as well as continuum opacity sources from Rayleigh scattering and collision-induced absorption of H$_2$ and He. Following the cloud prescription of \citet{2017MNRAS.469.1979M}, we assumed that the planetary atmosphere consisted of two equivalent components, a cloudless sector and a cloudy sector with a cloud fraction of $\phi$. In the cloudy sector, clouds and hazes were parameterized by the cloud top pressure $P_\mathrm{cloud}$ and the enhancement factor $A_\mathrm{RS}$ over nominal gas Rayleigh scattering, respectively. 

We used two approaches to treat the line opacities in addition to the filler gases H$_2$ and He. In the first approach, we assumed that the atmospheric species were in chemical equilibrium. We adopted the precalculated chemical equilibrium grid provided in the {\tt poor\_mans\_nonequ\_chem} subpackage of {\tt petitRADTRANS}, where the following 14 species are available for use: Na, K, TiO, VO, H$_2$O, FeH, CH$_4$, CO, CO$_2$, HCN, C$_2$H$_2$, H$_2$S, SiO, NH$_3$, PH$_3$. We interpolated the mass mixing ratio $X_i$ of the species $i$ in the grid using the carbon-to-oxygen number ratio ($\mathrm{C/O}$) and the metallicity $Z$ along with the temperature-pressure profile. 

In the second approach, we imposed no chemical constraints on the atmospheric species. For the list of species in the first approach, we kept only those with significant spectral signatures in the wavelength range covered by the OSIRIS and MuSCAT2 data. We also added a few metal hydrides with optical spectral band heads observable in the atmospheres of brown dwarfs \citep{2010ApJ...716.1060V}. This results in 10 species in use: Na, K, TiO, VO, H$_2$O, FeH, CaH, CrH, MgH, AlH. For each species, the mass mixing ratio $X_i$ was a free parameter and constant over the calculated pressure range, free from any chemical consideration. The filler gases H$_2$ and He had a mass mixing ratio of 3:1.

On the data side, we used a free offset $\Delta_\mathrm{GTC}$ for the OSIRIS spectrum to eliminate possible systematic offsets introduced by different instruments. Uniform or log-uniform priors were adopted for all the free parameters, as listed in Table~\ref{tab:retrieved_atm_param}. To perform model comparisons for the spectral retrievals, we computed the Bayes factor $\mathcal{B}_\mathrm{ab} = \mathcal{Z}_\mathrm{a}/\mathcal{Z}_\mathrm{b}$ for the a and b models. Following \citet{2008ConPh..49...71T} and \citet{2013ApJ...778..153B}, we adopted the $\ln \mathcal{B}_\mathrm{ab}=\Delta\ln\mathcal{Z}$ values of 1.0, 2.5, 5.0 as starting points for ``weak'', ``moderate'', ``strong'' evidence in favor of the a-model over the b-model, and also converted the Bayes factor approximately to frequentist significance values. 

\begin{table}
\caption[]{Parameter estimation for spectral retrievals.}
\renewcommand\arraystretch{1.5} 
\begin{center}
\setlength{\tabcolsep}{1.5mm}
\scalebox{0.8}{
\begin{tabular}{cccc}
\hline\hline
Parameter & Prior  & \multicolumn{2}{c}{Posterior}\\
 &   & Equilibrium chemistry   & Free chemistry\\
\hline
$T_\mathrm{iso}$ (K) 
& ${\mathcal{U}}(500,2500)$ 
& $2440^{+43}_{-78}$ 
& $2324^{+124}_ {-235}$\\
$\log P_0$ (bar) 
& $\mathcal{U}(-6,2)$ 
& $-0.69^{+0.33}_{-0.33}$ 
& $-2.44^{+0.27}_{-0.25}$\\
$\log P_\mathrm{cloud}$ (bar) 
& $\mathcal{U}(-6,2)$ 
& $0.40^{+1.04}_{-1.07}$ 
& $-0.53^{+1.65}_{-1.62}$\\
$\log A_\mathrm{RS}$ 
& $\mathcal{U}(-2,4)$ 
& $-0.43^{+1.51}_{-0.99}$ 
& $0.11^{+1.41}_{-1.33}$\\
$\phi$ 
& $\mathcal{U}(0,1)$ 
& $0.36^{+0.39}_{-0.29}$
& $0.44^{+0.36}_{-0.31}$\\
$\Delta_\mathrm{GTC}$ (ppm) 
& $\mathcal{U}(-5000,5000)$ 
& $-249^{+127}_{-131}$ 
& $-366^{+126}_{-119}$\\
$\log Z$ ($Z_\odot$)
& $\mathcal{U}(-2,3)$ 
& $-0.46^{+0.54}_{-0.60}$ 
& --\\
$\mathrm{C/O}$ 
& $\mathcal{U}(0.1,1.6)$ 
& $1.38^{+0.15}_{-0.19}$ 
& --\\
$\log X_\mathrm{Na}$ 
& $\mathcal{U}(-10,0)$  
& -- 
& $-1.22^{+0.32}_{-0.42}$\\
$\log X_\mathrm{K}$ 
& $\mathcal{U}(-10,0)$  
& -- 
& $-7.39^{+1.82}_{-1.69}$\\
$\log X_\mathrm{TiO}$ 
& $\mathcal{U}(-10,0)$  
& -- 
& $-8.91^{+0.80}_{-0.71}$\\
$\log X_\mathrm{VO}$ 
& $\mathcal{U}(-10,0)$  
& -- 
& $-8.71^{+0.94}_{-0.84}$\\
$\log X_\mathrm{H_2O}$ 
& $\mathcal{U}(-10,0)$  
& -- 
& $-5.86^{+2.76}_{-2.63}$\\
$\log X_\mathrm{FeH}$ 
& $\mathcal{U}(-10,0)$  
& -- 
& $-6.85^{+1.88}_{-2.04}$\\
$\log X_\mathrm{CaH}$ 
& $\mathcal{U}(-10,0)$  
& -- 
& $-9.62^{+0.40}_{-0.26}$\\
$\log X_\mathrm{CrH}$ 
& $\mathcal{U}(-10,0)$  
& -- 
& $-7.10^{+1.52}_{-1.79}$\\
$\log X_\mathrm{MgH}$ 
& $\mathcal{U}(-10,0)$  
& -- 
& $-4.00^{+0.87}_{-1.18}$\\
$\log X_\mathrm{AlH}$ 
& $\mathcal{U}(-10,0)$  
& -- 
& $-5.32^{+2.75}_{-3.07}$\\
\hline
\end{tabular}
}
\label{tab:retrieved_atm_param}  
\end{center}
\end{table}

\begin{table}
\caption[]{Bayesian model comparison for the free chemistry retrievals.} 
\renewcommand\arraystretch{1.5} 
\begin{center}
\setlength{\tabcolsep}{1.5mm}
\scalebox{0.85}{
\begin{tabular}{cccc} 
\hline\hline 
Species & $\Delta\ln\mathcal{Z}$\tablefootmark{a} & Significance & Inferences \\ 
\hline 
Na  & $13.48$ & $5.5\sigma$ & Strongly favored \\
K   & $-0.39$ & -- & Inconclusive  \\
TiO & $-1.46$ & $-2.3\sigma$ & Weakly disfavored  \\
VO  & $-1.38$ & $-2.2\sigma$ & Weakly disfavored  \\
H$_2$O & $-0.27$ & -- & Inconclusive \\
FeH & $-0.55$ & -- & Inconclusive \\
CaH & $-2.96$ & $-2.9\sigma$ & Moderately disfavored \\
CrH & $-0.47$ & -- & Inconclusive \\
MgH & $4.43$ & $3.4\sigma$ & Moderately favored \\
AlH & $-0.08$ & -- & Inconclusive \\
\hline 
\end{tabular}
}
\tablefoot{
\tablefoottext{a}{$\Delta\ln\mathcal{Z}=\ln\mathcal{Z}_\mathrm{full}-\ln\mathcal{Z}_\mathrm{i}$, where $\ln\mathcal{Z}_\mathrm{full}$ is the natural logarithmic evidence of the full model including all species and $\ln\mathcal{Z}_\mathrm{i}$ is the natural logarithmic evidence of the model excluding species $i$. The typical uncertainty of $\ln\mathcal{Z}$ in this study is $\sim$0.13.}
}
\label{tab:evidence} 
\end{center}
\end{table}

\subsubsection{Retrieval results}
\label{sec:retrieval_results}
\begin{figure*}[ht!]
\centering
\includegraphics[width=1.0\textwidth]{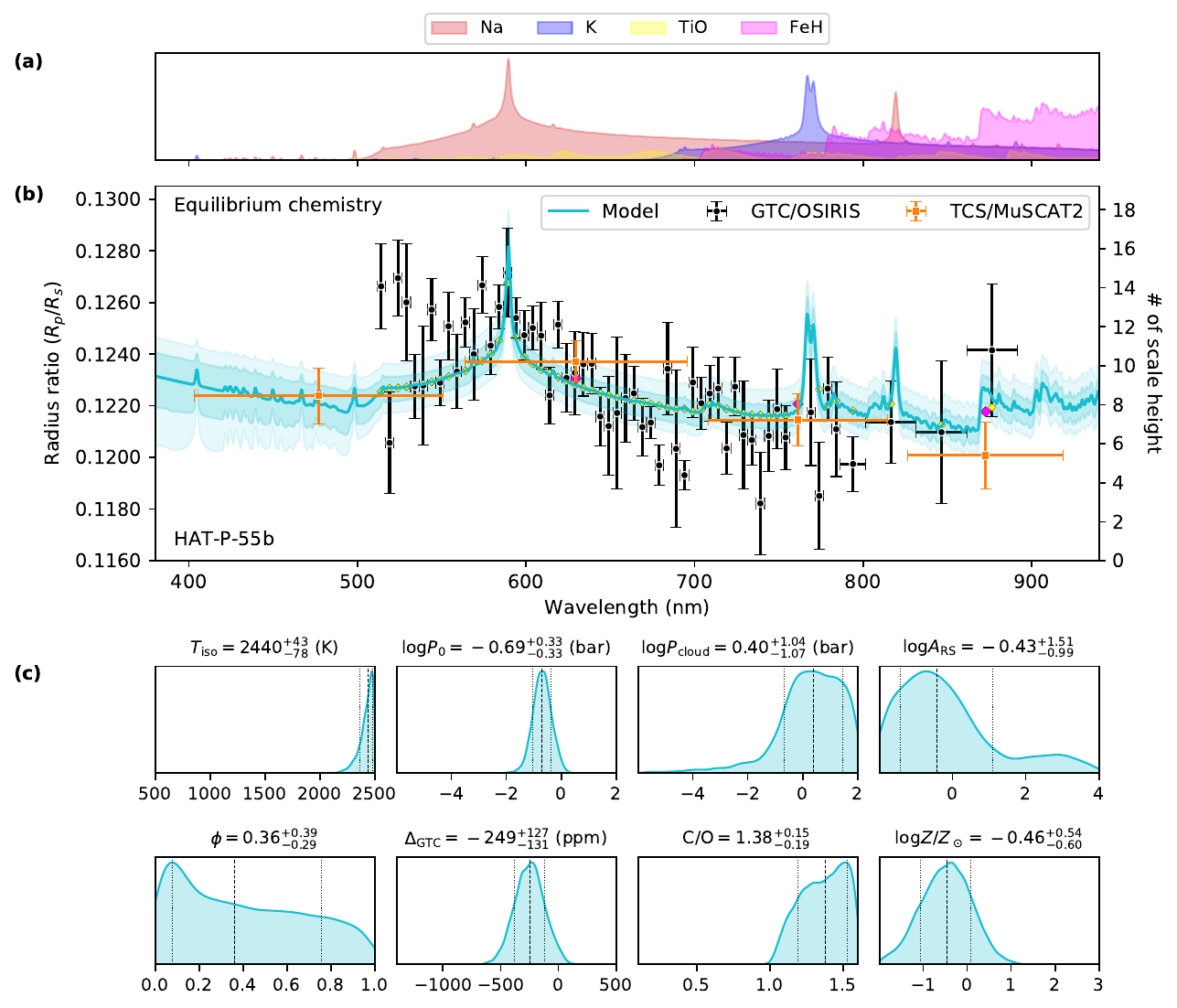}
\caption{Retrieval analysis of the transmission spectra of HAT-P-55b assuming equilibrium chemistry. Panel (a) shows the main species contributing to the spectral signatures. Panel (b) shows the transmission spectra obtained with GTC/OSIRIS and TCS/MuSCAT2, along with the median, 1-/2-$\sigma$ confidence levels of the retrieved models. Panel (c) shows the posterior distributions of the free parameters in the assumed atmospheric model.}
\label{fig: retrieval_eqchem}
\end{figure*}

\begin{figure*}[ht!]
\centering
\includegraphics[width=1.0\textwidth]{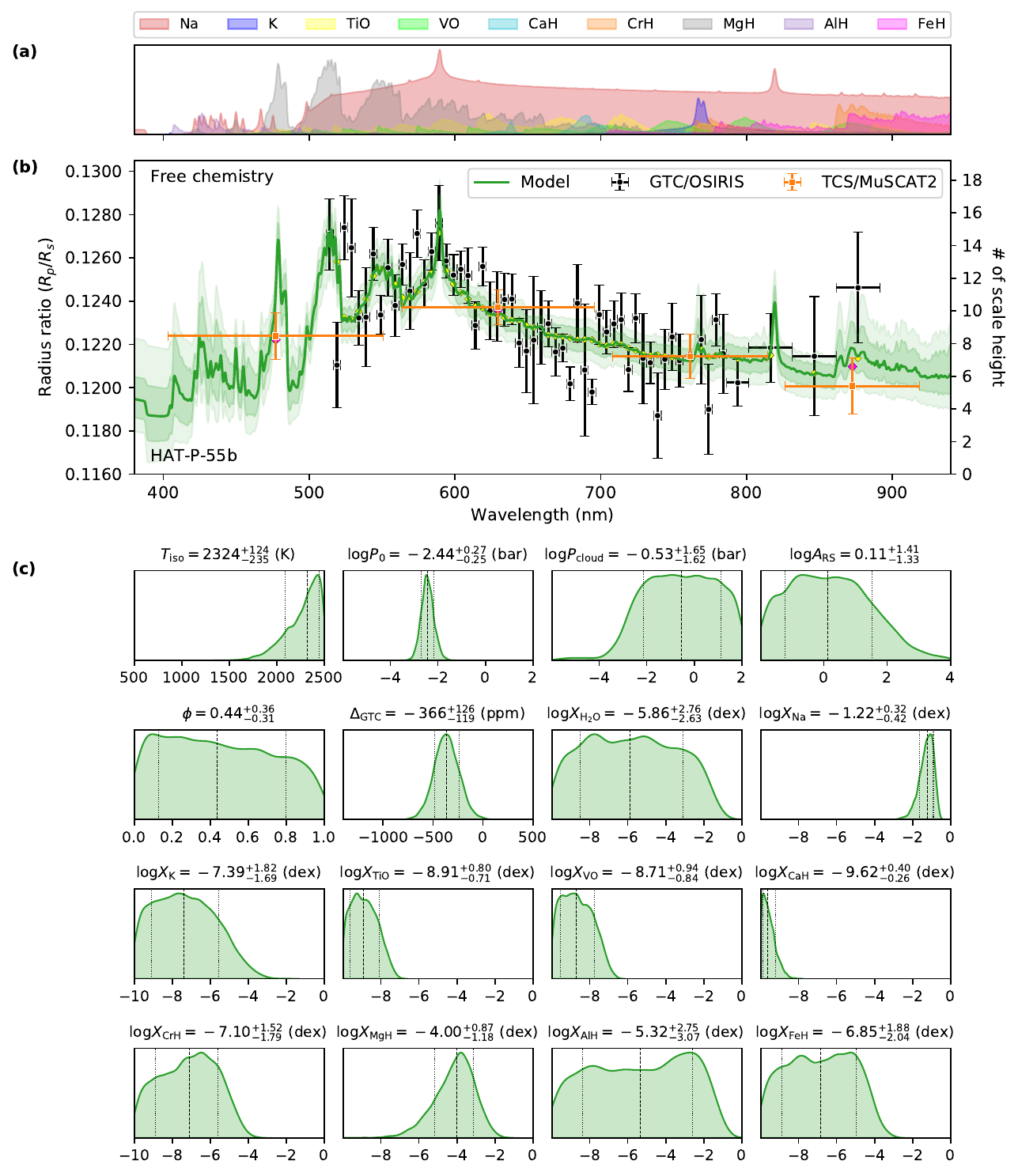}
\caption{Retrieval analysis of the transmission spectra of HAT-P-55b assuming free chemistry. Panel (a) shows the main species contributing to the spectral signatures. Panel (b) shows the transmission spectra obtained with GTC/OSIRIS and TCS/MuSCAT2, along with the median, 1-/2-$\sigma$ confidence levels of the retrieved models. Panel (c) shows the posterior distributions of the free parameters in the assumed atmospheric model.}
\label{fig: retrieval_freechem}
\end{figure*}

The retrieved models and parameters of our two assumptions are shown in Fig.~\ref{fig: retrieval_eqchem}, Fig.~\ref{fig: retrieval_freechem}, and Table~\ref{tab:retrieved_atm_param}. Table~\ref{tab:evidence} presents the Bayesian model comparison for the retrievals assuming free chemistry. 

Our retrieval analyses begin with the assumption of equilibrium chemistry. The combined OSIRIS and MuSCAT2 data can tightly constrain the reference pressure to $0.20^{+0.23}_{-0.11}$~bar and the instrumental offset to $-249^{+127}_{-131}$~ppm. The retrieved cloud top pressure of $2.5^{+25.2}_{-2.3}$~bar indicates that the clouds are relatively deep in the cloudy sector, while the scattering enhancement of $0.37^{+11.82}_{-0.34}$ is consistent with gas Rayleigh scattering, which together could cause the unconstrained cloud fraction of $36^{+39}_{-29}$\%. The retrieved atmospheric temperature $2440^{+43}_{-78}$~K is strongly skewed toward the upper boundary of the prior, much higher than the equilibrium temperature $1367\pm 11$~K \citep{2024MNRAS.528.1930K}. Our equilibrium chemistry retrieval also shows that the atmosphere of HAT-P-55b has a metallicity of $0.35^{+0.87}_{-0.26}\times$ solar and a high C/O ratio of $1.38^{+0.15}_{-0.19}$. Accordingly, the oxide molecules such as H$_2$O, TiO, and VO are deficient, and Na absorption with a prominent pressure broadening is the dominant opacity source.

Our free chemistry retrievals are used to explore which species might be responsible for the observed spectral signatures. The full free chemistry model yielded an atmospheric temperature of $2324^{+124}_{-235}$~K, a cloud top pressure of $0.29^{+12.7}_{-0.29}$~bar, a scattering enhancement of $1.3^{+31.8}_{-1.2}$, and a cloud fraction of $44^{+36}_{-31}$\%, consistent with those in the equilibrium chemistry assumption. The only significant deviation comes from the retrieved reference pressure, which drops to $3.6^{+3.1}_{-1.6}$~mbar in the free chemistry assumption. Of the 10 species included, only the mass mixing ratios of Na and MgH are constrained, $-1.22^{+0.32}_{-0.42}$ and $-4.00^{+0.87}_{-1.18}$ dex, while those of H$_2$O and AlH are unconstrained. For the rest, 3$\sigma$ upper limits of $<$$-3.11$ dex are obtained for K, $<$$-7.11$ dex for TiO, $<$$-6.66$ dex for VO, $<$$-3.27$ for FeH, $<$$-8.22$ dex for CaH, and $<$$-3.90$ dex for CrH. 

However, these mass mixing ratios in the free chemistry retrieval, as well as the metallicity in the equilibrium chemistry retrieval, may be biased because the current wavelengths covered by OSIRIS and MuSCAT2 are limited to optical, which has poor coverage of the major molecules contributing to metallicity and C/O (e.g., H$_2$O, CO, CO$_2$, CH$_4$). The relative difference between optical absorbers and infrared molecules is crucial for constraining reference pressure and chemical abundances \citep{2019AJ....157..206W}. Indeed, strong anticorrelations are found between reference pressure and metallicity, between reference pressure and mass mixing ratios, and between reference pressure and instrumental offset. On the other hand, the mass mixing ratios of Na and MgH are strongly correlated. Given these degeneracies, which could only be mitigated by future follow-up observations in the infrared, relative abundances are more reliable. The relative mass mixing ratio between MgH and Na measured in the free chemistry retrieval, $\log X_\mathrm{MgH}/X_\mathrm{Na}=-2.82^{+0.95}_{-0.84}$~dex, is consistent with the prediction of equilibrium chemistry with solar composition ($-2.80$~dex). 

The free chemistry retrieval was repeated after removing each of the 10 species, and the model evidence was compared to that of the full free chemistry model to assess the contribution of that species. As a result, Na is strongly favored with $\Delta\ln\mathcal{Z}=13.48$ (5.5$\sigma$) and MgH is moderately favored with $\Delta\ln\mathcal{Z}=4.43$ (3.4$\sigma$). In contrast, CaH is moderately disfavored ($\Delta\ln\mathcal{Z}=-2.96$, 2.9$\sigma$), while TiO ($\Delta\ln\mathcal{Z}=-1.46$, 2.3$\sigma$) and VO ($\Delta\ln\mathcal{Z}=-1.38$, 2.2$\sigma$) are weakly disfavored.

To confirm whether the combinations of other species could mimic the observed broad spectral profile provided by Na, we performed additional retrieval tests on a more extensive list of species with potential spectral signatures within 510--700~nm, including Na, K, TiO, VO, H$_2$O, FeH, CaH, CrH, MgH, AlH, H$_2$S, HCN, NH$_3$, AlO, Fe, Li, Ti, V. The results still strongly favor the presence of Na with $\Delta\ln\mathcal{Z}=8.80$ (4.6$\sigma$), ruling out the possibility of compensating for the absence of Na absorption by combinations of other species. 

To assess whether the Na detection was driven by specific data points, we also performed additional retrieval tests by masking some passbands of the OSIRIS transmission spectrum. Masking the 589~nm passband would lead to Na detection with $\Delta\ln\mathcal{Z}=11.79$ (5.2$\sigma$), while masking three data points (584~nm, 589~nm, 594~nm) reduces the significance to $\Delta\ln\mathcal{Z}=7.06$ (4.2$\sigma$). This indicates that the current inference of Na detection is driven not only by the few data points at the line core of Na, but also by its very broad line wing. 

\section{Conclusions and discussion}
\label{sec:conclusions}
We have conducted a follow-up spectrophotometric transit observation of the hot Jupiter HAT-P-55b using GTC/OSIRIS. We derived transmission spectra for this planet using the conventional band-integrated method and the new pixel-based method, which are consistent with each other and in broad agreement with the tentative color signature previously observed by \citet{2024MNRAS.528.1930K} using TCS/MuSCAT2. We performed Bayesian spectral retrieval analyses on the combined OSIRIS and MuSCAT2 transmission spectrum and found that the observed spectral signature is likely due to Na and possibly partially from MgH. While it is difficult to constrain the absolute abundances of Na and MgH with the current optical-only wavelength coverage, the relative mass mixing ratio between Na and MgH is consistent with the prediction of the solar composition at chemical equilibrium. 

\begin{figure}
\centering
\includegraphics[width=1.0\linewidth]{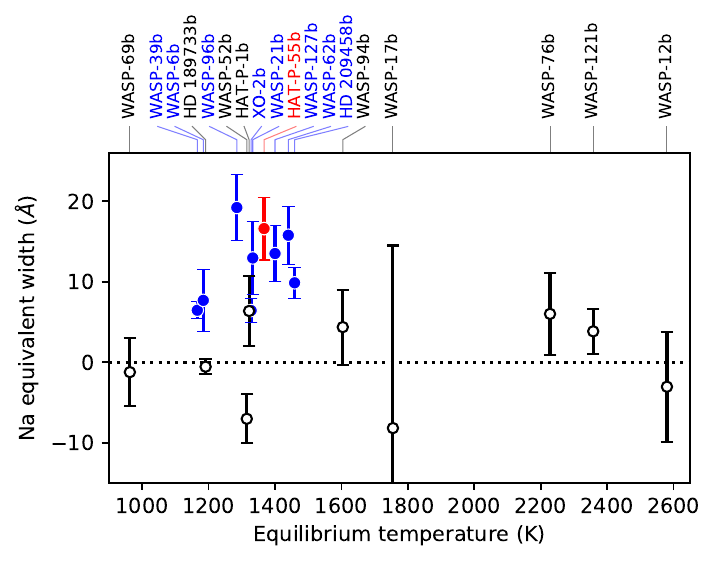}
\caption{Equivalent width of Na measured in low-resolution transmission spectra for planets with Na detection reported. Measurements with S/N greater than 2 are shown as filled circles, while the others are shown as empty circles. HAT-P-55b is highlighted in red.}
\label{fig: Na_EW}
\end{figure}

\begin{table}
\caption[]{Equivalent widths of Na in low-resolution transmission spectra for planets with Na detection reported.} 
\renewcommand\arraystretch{1.5} 
\begin{center}
\setlength{\tabcolsep}{1.5mm}
\scalebox{0.85}{
\begin{tabular}{cccc} 
\hline\hline 
Planet & $T_\mathrm{eq}$ (K)\tablefootmark{a} & $EW_\mathrm{Na}$ ($\AA$)\tablefootmark{b} & Reference \\ 
\hline 
HAT-P-1b   & $1322$ & $  6.4 \pm  4.3$ & 1, 2\\
HAT-P-55b  & $1367$ & $ 16.6 \pm  3.9$ & 3\\
HD 189733b & $1191$ & $ -0.6 \pm  0.9$ & 1\\
HD 209458b & $1459$ & $  9.9 \pm  1.9$ & 1\\
WASP-6b    & $1184$ & $  1.7 \pm  2.8$ & 4\\
WASP-12b   & $2580$ & $ -3.0 \pm  6.8$ & 1\\
WASP-17b   & $1755$ & $ -8.2 \pm 22.7$ & 5\\
WASP-21b   & $1333$ & $ 12.9 \pm  4.6$ & 6, 7\\
WASP-39b   & $1166$ & $  6.5 \pm  1.0$ & 1, 8\\
WASP-52b   & $1315$ & $ -7.0 \pm  3.0$ & 9, 10\\
WASP-62b   & $1440$ & $ 15.8 \pm  3.6$ & 11\\
WASP-69b   & $ 963$ & $ -1.2 \pm  4.2$ & 12\\
WASP-76b   & $2228$ & $  6.0 \pm  5.1$ & 13\\
WASP-94b   & $1604$ & $  4.4 \pm  4.7$ & 14\\
WASP-96b   & $1285$ & $ 19.2 \pm  4.1$ & 15\\
WASP-121b  & $2358$ & $  3.8 \pm  2.8$ & 16, 17\\
WASP-127b  & $1400$ & $ 13.5 \pm  3.5$ & 18, 19\\
XO-2b      & $1328$ & $  6.4 \pm  1.5$ & 20\\
\hline 
\end{tabular}
}
\tablefoot{
\tablefoottext{a}{Equilibrium temperatures $T_\mathrm{eq}$ are taken from TEPCat \citep{2011MNRAS.417.2166S}.}
\tablefoottext{b}{For planets with multiple measurements, the weighted mean values are reported.}
}
\tablebib{
(1) \citet{2016Natur.529...59S}; 
(2) \citet{2022AJ....164..173C}; 
(3) This work; 
(4) \citet{2020MNRAS.494.5449C}; 
(5) \citet{2022MNRAS.512.4185A}; 
(6) \citet{2020A&A...642A..54C}; 
(7) \citet{2020MNRAS.497.5182A}; 
(8) \citet{2023Natur.614..659R}; 
(9) \citet{2017A&A...600L..11C}; 
(10) \citet{2018AJ....156..298A}; 
(11) \citet{2021ApJ...906L..10A}; 
(12) \citet{2020A&A...641A.158M}; 
(13) \citet{2021AJ....162..108F}; 
(14) \citet{2022MNRAS.510.4857A}; 
(15) \citet{2018Natur.557..526N};  
(16) \citet{2018AJ....156..283E};  
(17) \citet{2021MNRAS.503.4787W}; 
(18) \citet{2018A&A...616A.145C};  
(19) \citet{2021MNRAS.500.4042S}; 
(20) \citet{2019AJ....157...21P}.
}
\label{tab:Na_EW}
\end{center}
\end{table}

The prominent broad Na absorption signature makes HAT-P-55b a new addition to the small group of hot Jupiters that exhibit potential alkali pressure broadening. Following the line broadening metric proposed by \citet{2020A&A...642A..54C} but without normalization, we calculated the equivalent width within an 80~nm band centered on the Na line in low-resolution transmission spectra for planets with Na detection reported in the literature. As shown in Fig.~\ref{fig: Na_EW} and Table~\ref{tab:Na_EW}, nine planets have large positive  equivalent widths for Na with S/N greater than 2, indicating potentially pressure-broadened Na line profiles, including WASP-39b, WASP-6b, WASP-96b, XO-2b, WASP-21b, HAT-P-55b, WASP-127b, WASP-62b, and HD 209458b in order of equilibrium temperature. This group of hot Jupiters holds great promise for precise abundance measurements and well-determined cloud properties when combined with abundant molecular bands to be observed by the James Webb Space Telescope (JWST) and Hubble Space Telescope (HST). Such measurements could help answer the question of whether there are trends between alkali metal abundances and some stellar and planetary parameters \citep{2019ApJ...887L..20W,2023AJ....166..120M,2024arXiv240208292S}. 

On the other hand, the moderate evidence for the presence of MgH requires high-resolution spectroscopy to confirm \citep{2020AJ....160..228K,2023ApJ...953L..19F}. Metal hydrides have been proposed as probes of atmospheric chemistry, cloud formation and potential sources of opacity for thermal inversions \citep{2008ApJ...674..451B,2010ApJ...716.1060V,2018ApJ...866...27L}, but are difficult to detect in the atmospheres of hot Jupiters. Preliminary evidence for several metal hydrides (e.g., FeH, TiH, CrH, SiH, MgH) has been reported from low-resolution transmission spectra \citep{2016ApJ...822L...4E,2019MNRAS.486.1292M,2020AJ....159....5S,2020AJ....160..109S,2021ApJ...906L..10A,2021A&A...646A..17B,2024A&A...682A..73J}, but only one case has been confirmed by high-resolution transmission spectroscopy, i.e., CrH in WASP-31b \citep{2023ApJ...953L..19F}. 

We conclude that HAT-P-55b is a promising target for further detailed atmospheric characterization with high-quality observations from JWST, HST, and ground-based high-resolution spectrographs.

\begin{acknowledgements}
The authors thank the anonymous referee for their valuable comments and suggestions. 
G.C. acknowledges the support by the National Natural Science Foundation of China (NSFC grant Nos. 12122308, 42075122), the B-type Strategic Priority Program of the Chinese Academy of Sciences (grant No. XDB41000000), Youth Innovation Promotion Association CAS (2021315), and the Minor Planet Foundation of the Purple Mountain Observatory. 
Y.M. acknowledges the support by NSFC (grant No. 12033010). 
This work is partly supported by JSPS KAKENHI Grant Number JPJP24H00017 and JSPS Bilateral Program Number JPJSBP120249910. 
This work is based on observations made with the Gran Telescopio Canarias installed at the Spanish Observatorio del Roque de los Muchachos of the Instituto de Astrofísica de Canarias on the island of La Palma. 
\end{acknowledgements}

%
 
\bibliographystyle{aa.bst} 
\bibliography{ref.bib}

\begin{thebibliography}{70}
\expandafter\ifx\csname natexlab\endcsname\relax\def\natexlab#1{#1}\fi

\bibitem[{{Ahrer} {et~al.}(2022){Ahrer}, {Wheatley}, {Kirk}, {Gandhi}, {King},
  \& {Louden}}]{2022MNRAS.510.4857A}
{Ahrer}, E., {Wheatley}, P.~J., {Kirk}, J., {et~al.} 2022, \mnras, 510, 4857

\bibitem[{{Alam} {et~al.}(2021){Alam}, {L{\'o}pez-Morales}, {MacDonald},
  {Nikolov}, {Kirk}, {Goyal}, {Sing}, {Wakeford}, {Rathcke}, {Deming},
  {Sanz-Forcada}, {Lewis}, {Barstow}, {Mikal-Evans}, \&
  {Buchhave}}]{2021ApJ...906L..10A}
{Alam}, M.~K., {L{\'o}pez-Morales}, M., {MacDonald}, R.~J., {et~al.} 2021,
  \apjl, 906, L10

\bibitem[{{Alam} {et~al.}(2018){Alam}, {Nikolov}, {L{\'o}pez-Morales}, {Sing},
  {Goyal}, {Henry}, {Sanz-Forcada}, {Williamson}, {Evans}, {Wakeford}, {Bruno},
  {Ballester}, {Stevenson}, {Lewis}, {Barstow}, {Bourrier}, {Buchhave},
  {Ehrenreich}, \& {Garc{\'\i}a Mu{\~n}oz}}]{2018AJ....156..298A}
{Alam}, M.~K., {Nikolov}, N., {L{\'o}pez-Morales}, M., {et~al.} 2018, \aj, 156,
  298

\bibitem[{{Alderson} {et~al.}(2020){Alderson}, {Kirk}, {L{\'o}pez-Morales},
  {Wheatley}, {Skillen}, {Henry}, {McGruder}, {Brogi}, {Louden}, \&
  {King}}]{2020MNRAS.497.5182A}
{Alderson}, L., {Kirk}, J., {L{\'o}pez-Morales}, M., {et~al.} 2020, \mnras,
  497, 5182

\bibitem[{{Alderson} {et~al.}(2022){Alderson}, {Wakeford}, {MacDonald},
  {Lewis}, {May}, {Grant}, {Sing}, {Stevenson}, {Fowler}, {Goyal}, {Batalha},
  \& {Kataria}}]{2022MNRAS.512.4185A}
{Alderson}, L., {Wakeford}, H.~R., {MacDonald}, R.~J., {et~al.} 2022, \mnras,
  512, 4185

\bibitem[{{Ambikasaran} {et~al.}(2015){Ambikasaran}, {Foreman-Mackey},
  {Greengard}, {Hogg}, \& {O'Neil}}]{2015ITPAM..38..252A}
{Ambikasaran}, S., {Foreman-Mackey}, D., {Greengard}, L., {Hogg}, D.~W., \&
  {O'Neil}, M. 2015, IEEE Transactions on Pattern Analysis and Machine
  Intelligence, 38, 252

\bibitem[{{Benneke} \& {Seager}(2012)}]{2012ApJ...753..100B}
{Benneke}, B. \& {Seager}, S. 2012, \apj, 753, 100

\bibitem[{{Benneke} \& {Seager}(2013)}]{2013ApJ...778..153B}
{Benneke}, B. \& {Seager}, S. 2013, \apj, 778, 153

\bibitem[{{B{\'e}tr{\'e}mieux} \& {Swain}(2017)}]{2017MNRAS.467.2834B}
{B{\'e}tr{\'e}mieux}, Y. \& {Swain}, M.~R. 2017, \mnras, 467, 2834

\bibitem[{{Braam} {et~al.}(2021){Braam}, {van der Tak}, {Chubb}, \&
  {Min}}]{2021A&A...646A..17B}
{Braam}, M., {van der Tak}, F. F.~S., {Chubb}, K.~L., \& {Min}, M. 2021, \aap,
  646, A17

\bibitem[{{Buchner} {et~al.}(2014){Buchner}, {Georgakakis}, {Nandra}, {Hsu},
  {Rangel}, {Brightman}, {Merloni}, {Salvato}, {Donley}, \&
  {Kocevski}}]{2014A&A...564A.125B}
{Buchner}, J., {Georgakakis}, A., {Nandra}, K., {et~al.} 2014, \aap, 564, A125

\bibitem[{{Burgasser} {et~al.}(2008){Burgasser}, {Looper}, {Kirkpatrick},
  {Cruz}, \& {Swift}}]{2008ApJ...674..451B}
{Burgasser}, A.~J., {Looper}, D.~L., {Kirkpatrick}, J.~D., {Cruz}, K.~L., \&
  {Swift}, B.~J. 2008, \apj, 674, 451

\bibitem[{{Carter} {et~al.}(2020){Carter}, {Nikolov}, {Sing}, {Alam}, {Goyal},
  {Mikal-Evans}, {Wakeford}, {Henry}, {Morrell}, {L{\'o}pez-Morales},
  {Smalley}, {Lavvas}, {Barstow}, {Garc{\'\i}a Mu{\~n}oz}, {Gibson}, \&
  {Wilson}}]{2020MNRAS.494.5449C}
{Carter}, A.~L., {Nikolov}, N., {Sing}, D.~K., {et~al.} 2020, \mnras, 494, 5449

\bibitem[{{Cepa} {et~al.}(2000){Cepa}, {Aguiar}, {Escalera},
  {Gonzalez-Serrano}, {Joven-Alvarez}, {Peraza}, {Rasilla}, {Rodriguez-Ramos},
  {Gonzalez}, {Cobos Duenas}, {Sanchez}, {Tejada}, {Bland-Hawthorn},
  {Militello}, \& {Rosa}}]{2000SPIE.4008..623C}
{Cepa}, J., {Aguiar}, M., {Escalera}, V.~G., {et~al.} 2000, in Society of
  Photo-Optical Instrumentation Engineers (SPIE) Conference Series, Vol. 4008,
  Optical and IR Telescope Instrumentation and Detectors, ed. M.~{Iye} \& A.~F.
  {Moorwood}, 623--631

\bibitem[{{Chen} {et~al.}(2020){Chen}, {Casasayas-Barris}, {Pall{\'e}},
  {Welbanks}, {Madhusudhan}, {Luque}, \& {Murgas}}]{2020A&A...642A..54C}
{Chen}, G., {Casasayas-Barris}, N., {Pall{\'e}}, E., {et~al.} 2020, \aap, 642,
  A54

\bibitem[{{Chen} {et~al.}(2017{\natexlab{a}}){Chen}, {Guenther}, {Pall{\'e}},
  {Nortmann}, {Nowak}, {Kunz}, {Parviainen}, \& {Murgas}}]{2017A&A...600A.138C}
{Chen}, G., {Guenther}, E.~W., {Pall{\'e}}, E., {et~al.} 2017{\natexlab{a}},
  \aap, 600, A138

\bibitem[{{Chen} {et~al.}(2017{\natexlab{b}}){Chen}, {Pall{\'e}}, {Nortmann},
  {Murgas}, {Parviainen}, \& {Nowak}}]{2017A&A...600L..11C}
{Chen}, G., {Pall{\'e}}, E., {Nortmann}, L., {et~al.} 2017{\natexlab{b}}, \aap,
  600, L11

\bibitem[{{Chen} {et~al.}(2018){Chen}, {Pall{\'e}}, {Welbanks},
  {Prieto-Arranz}, {Madhusudhan}, {Gandhi}, {Casasayas-Barris}, {Murgas},
  {Nortmann}, {Crouzet}, {Parviainen}, \& {Gandolfi}}]{2018A&A...616A.145C}
{Chen}, G., {Pall{\'e}}, E., {Welbanks}, L., {et~al.} 2018, \aap, 616, A145

\bibitem[{{Chen} {et~al.}(2022){Chen}, {Wang}, {van Boekel}, \&
  {Pall{\'e}}}]{2022AJ....164..173C}
{Chen}, G., {Wang}, H., {van Boekel}, R., \& {Pall{\'e}}, E. 2022, \aj, 164,
  173

\bibitem[{{de Wit} \& {Seager}(2013)}]{2013Sci...342.1473D}
{de Wit}, J. \& {Seager}, S. 2013, Science, 342, 1473

\bibitem[{{Eastman} {et~al.}(2010){Eastman}, {Siverd}, \&
  {Gaudi}}]{2010PASP..122..935E}
{Eastman}, J., {Siverd}, R., \& {Gaudi}, B.~S. 2010, \pasp, 122, 935

\bibitem[{{Espinoza} \& {Jord{\'a}n}(2015)}]{2015MNRAS.450.1879E}
{Espinoza}, N. \& {Jord{\'a}n}, A. 2015, \mnras, 450, 1879

\bibitem[{{Evans} {et~al.}(2018){Evans}, {Sing}, {Goyal}, {Nikolov}, {Marley},
  {Zahnle}, {Henry}, {Barstow}, {Alam}, {Sanz-Forcada}, {Kataria}, {Lewis},
  {Lavvas}, {Ballester}, {Ben-Jaffel}, {Blumenthal}, {Bourrier}, {Drummond},
  {Garc{\'\i}a Mu{\~n}oz}, {L{\'o}pez-Morales}, {Tremblin}, {Ehrenreich},
  {Wakeford}, {Buchhave}, {Lecavelier des Etangs}, {H{\'e}brard}, \&
  {Williamson}}]{2018AJ....156..283E}
{Evans}, T.~M., {Sing}, D.~K., {Goyal}, J.~M., {et~al.} 2018, \aj, 156, 283

\bibitem[{{Evans} {et~al.}(2016){Evans}, {Sing}, {Wakeford}, {Nikolov},
  {Ballester}, {Drummond}, {Kataria}, {Gibson}, {Amundsen}, \&
  {Spake}}]{2016ApJ...822L...4E}
{Evans}, T.~M., {Sing}, D.~K., {Wakeford}, H.~R., {et~al.} 2016, \apjl, 822, L4

\bibitem[{{Feinstein} {et~al.}(2023){Feinstein}, {Radica}, {Welbanks},
  {Murray}, {Ohno}, {Coulombe}, {Espinoza}, {Bean}, {Teske}, {Benneke}, {Line},
  {Rustamkulov}, {Saba}, {Tsiaras}, {Barstow}, {Fortney}, {Gao}, {Knutson},
  {MacDonald}, {Mikal-Evans}, {Rackham}, {Taylor}, {Parmentier}, {Batalha},
  {Berta-Thompson}, {Carter}, {Changeat}, {dos Santos}, {Gibson}, {Goyal},
  {Kreidberg}, {L{\'o}pez-Morales}, {Lothringer}, {Miguel}, {Molaverdikhani},
  {Moran}, {Morello}, {Mukherjee}, {Sing}, {Stevenson}, {Wakeford}, {Ahrer},
  {Alam}, {Alderson}, {Allen}, {Batalha}, {Bell}, {Blecic}, {Brande},
  {Caceres}, {Casewell}, {Chubb}, {Crossfield}, {Crouzet}, {Cubillos}, {Decin},
  {D{\'e}sert}, {Harrington}, {Heng}, {Henning}, {Iro}, {Kempton}, {Kendrew},
  {Kirk}, {Krick}, {Lagage}, {Lendl}, {Mancini}, {Mansfield}, {May}, {Mayne},
  {Nikolov}, {Palle}, {Petit dit de la Roche}, {Piaulet}, {Powell}, {Redfield},
  {Rogers}, {Roman}, {Roy}, {Nixon}, {Schlawin}, {Tan}, {Tremblin}, {Turner},
  {Venot}, {Waalkes}, {Wheatley}, \& {Zhang}}]{2023Natur.614..670F}
{Feinstein}, A.~D., {Radica}, M., {Welbanks}, L., {et~al.} 2023, \nat, 614, 670

\bibitem[{{Flagg} {et~al.}(2023){Flagg}, {Turner}, {Deibert}, {Ridden-Harper},
  {de Mooij}, {MacDonald}, {Jayawardhana}, {Gibson}, {Langeveld}, \&
  {Sing}}]{2023ApJ...953L..19F}
{Flagg}, L., {Turner}, J.~D., {Deibert}, E., {et~al.} 2023, \apjl, 953, L19

\bibitem[{{Foreman-Mackey} {et~al.}(2017){Foreman-Mackey}, {Agol},
  {Ambikasaran}, \& {Angus}}]{2017AJ....154..220F}
{Foreman-Mackey}, D., {Agol}, E., {Ambikasaran}, S., \& {Angus}, R. 2017, \aj,
  154, 220

\bibitem[{{Foreman-Mackey} {et~al.}(2013){Foreman-Mackey}, {Hogg}, {Lang}, \&
  {Goodman}}]{2013PASP..125..306F}
{Foreman-Mackey}, D., {Hogg}, D.~W., {Lang}, D., \& {Goodman}, J. 2013, \pasp,
  125, 306

\bibitem[{{Fortney} {et~al.}(2010){Fortney}, {Shabram}, {Showman}, {Lian},
  {Freedman}, {Marley}, \& {Lewis}}]{2010ApJ...709.1396F}
{Fortney}, J.~J., {Shabram}, M., {Showman}, A.~P., {et~al.} 2010, \apj, 709,
  1396

\bibitem[{{Fu} {et~al.}(2021){Fu}, {Deming}, {Lothringer}, {Nikolov}, {Sing},
  {Kempton}, {Ih}, {Evans}, {Stevenson}, {Wakeford}, {Rodriguez}, {Eastman},
  {Stassun}, {Henry}, {L{\'o}pez-Morales}, {Lendl}, {Conti}, {Stockdale},
  {Collins}, {Kielkopf}, {Barstow}, {Sanz-Forcada}, {Ehrenreich}, {Bourrier},
  \& {dos Santos}}]{2021AJ....162..108F}
{Fu}, G., {Deming}, D., {Lothringer}, J., {et~al.} 2021, \aj, 162, 108

\bibitem[{{Gibson} {et~al.}(2012){Gibson}, {Aigrain}, {Roberts}, {Evans},
  {Osborne}, \& {Pont}}]{2012MNRAS.419.2683G}
{Gibson}, N.~P., {Aigrain}, S., {Roberts}, S., {et~al.} 2012, \mnras, 419, 2683

\bibitem[{{Griffith}(2014)}]{2014RSPTA.37230086G}
{Griffith}, C.~A. 2014, Philosophical Transactions of the Royal Society of
  London Series A, 372, 20130086

\bibitem[{{Heng} \& {Kitzmann}(2017)}]{2017MNRAS.470.2972H}
{Heng}, K. \& {Kitzmann}, D. 2017, \mnras, 470, 2972

\bibitem[{{Horne}(1986)}]{1986PASP...98..609H}
{Horne}, K. 1986, \pasp, 98, 609

\bibitem[{{Jiang} {et~al.}(2024){Jiang}, {Chen}, {Murgas}, {Pall{\'e}},
  {Parviainen}, \& {Ma}}]{2024A&A...682A..73J}
{Jiang}, C., {Chen}, G., {Murgas}, F., {et~al.} 2024, \aap, 682, A73

\bibitem[{{Jiang} {et~al.}(2023){Jiang}, {Chen}, {Pall{\'e}}, {Murgas},
  {Parviainen}, \& {Ma}}]{2023A&A...675A..62J}
{Jiang}, C., {Chen}, G., {Pall{\'e}}, E., {et~al.} 2023, \aap, 675, A62

\bibitem[{{Jiang} {et~al.}(2022){Jiang}, {Chen}, {Pall{\'e}}, {Parviainen},
  {Murgas}, \& {Ma}}]{2022A&A...664A..50J}
{Jiang}, C., {Chen}, G., {Pall{\'e}}, E., {et~al.} 2022, \aap, 664, A50

\bibitem[{{Juncher} {et~al.}(2015){Juncher}, {Buchhave}, {Hartman}, {Bakos},
  {Bieryla}, {Kov{\'a}cs}, {Boisse}, {Latham}, {Kov{\'a}cs}, {Bhatti},
  {Csubry}, {Penev}, {de Val-Borro}, {Falco}, {Torres}, {Noyes},
  {L{\'a}z{\'a}r}, {Papp}, \& {S{\'a}ri}}]{2015PASP..127..851J}
{Juncher}, D., {Buchhave}, L.~A., {Hartman}, J.~D., {et~al.} 2015, \pasp, 127,
  851

\bibitem[{{Kang} {et~al.}(2024){Kang}, {Chen}, {Pall{\'e}}, {Murgas}, {Abreu
  Garc{\'\i}a}, {de Leon}, {Enoc}, {Esparza-Borges}, {Fukuda}, {Fukui},
  {Gal{\'a}n}, {Hayashi}, {Isogai}, {Kagetani}, {Kawauchi}, {Korth},
  {Livingston}, {Luque}, {Ma}, {Madrigal-Aguado}, {Meni}, {Monta{\~n}es
  Rodriguez}, {Mori}, {Mu{\~n}oz Torres}, {Narita}, {Orell-Miquel},
  {Parviainen}, {Pel{\'a}ez-Torres}, {Stangret}, {Tamura}, \&
  {Watanabe}}]{2024MNRAS.528.1930K}
{Kang}, H., {Chen}, G., {Pall{\'e}}, E., {et~al.} 2024, \mnras, 528, 1930

\bibitem[{{Kesseli} {et~al.}(2020){Kesseli}, {Snellen}, {Alonso-Floriano},
  {Molli{\`e}re}, \& {Serindag}}]{2020AJ....160..228K}
{Kesseli}, A.~Y., {Snellen}, I.~A.~G., {Alonso-Floriano}, F.~J.,
  {Molli{\`e}re}, P., \& {Serindag}, D.~B. 2020, \aj, 160, 228

\bibitem[{{Kreidberg}(2015)}]{2015PASP..127.1161K}
{Kreidberg}, L. 2015, \pasp, 127, 1161

\bibitem[{{Lecavelier Des Etangs} {et~al.}(2008){Lecavelier Des Etangs},
  {Pont}, {Vidal-Madjar}, \& {Sing}}]{2008A&A...481L..83L}
{Lecavelier Des Etangs}, A., {Pont}, F., {Vidal-Madjar}, A., \& {Sing}, D.
  2008, \aap, 481, L83

\bibitem[{{Lothringer} {et~al.}(2018){Lothringer}, {Barman}, \&
  {Koskinen}}]{2018ApJ...866...27L}
{Lothringer}, J.~D., {Barman}, T., \& {Koskinen}, T. 2018, \apj, 866, 27

\bibitem[{{Lothringer} {et~al.}(2021){Lothringer}, {Rustamkulov}, {Sing},
  {Gibson}, {Wilson}, \& {Schlaufman}}]{2021ApJ...914...12L}
{Lothringer}, J.~D., {Rustamkulov}, Z., {Sing}, D.~K., {et~al.} 2021, \apj,
  914, 12

\bibitem[{{MacDonald} \& {Madhusudhan}(2017)}]{2017MNRAS.469.1979M}
{MacDonald}, R.~J. \& {Madhusudhan}, N. 2017, \mnras, 469, 1979

\bibitem[{{MacDonald} \& {Madhusudhan}(2019)}]{2019MNRAS.486.1292M}
{MacDonald}, R.~J. \& {Madhusudhan}, N. 2019, \mnras, 486, 1292

\bibitem[{{Madhusudhan} {et~al.}(2014){Madhusudhan}, {Amin}, \&
  {Kennedy}}]{2014ApJ...794L..12M}
{Madhusudhan}, N., {Amin}, M.~A., \& {Kennedy}, G.~M. 2014, \apjl, 794, L12

\bibitem[{{Madhusudhan} \& {Seager}(2009)}]{2009ApJ...707...24M}
{Madhusudhan}, N. \& {Seager}, S. 2009, \apj, 707, 24

\bibitem[{{McGruder} {et~al.}(2023){McGruder}, {L{\'o}pez-Morales}, {Kirk},
  {Rackham}, {May}, {Ahrer}, {King}, {Alam}, {Allen}, {Ceballos}, {Espinoza},
  {Gardner}, {Jord{\'a}n}, {Meyer}, {Monnier}, {Osip}, \&
  {Wheatley}}]{2023AJ....166..120M}
{McGruder}, C.~D., {L{\'o}pez-Morales}, M., {Kirk}, J., {et~al.} 2023, \aj,
  166, 120

\bibitem[{{Molli{\`e}re} {et~al.}(2019){Molli{\`e}re}, {Wardenier}, {van
  Boekel}, {Henning}, {Molaverdikhani}, \& {Snellen}}]{2019A&A...627A..67M}
{Molli{\`e}re}, P., {Wardenier}, J.~P., {van Boekel}, R., {et~al.} 2019, \aap,
  627, A67

\bibitem[{{Mordasini} {et~al.}(2016){Mordasini}, {van Boekel}, {Molli{\`e}re},
  {Henning}, \& {Benneke}}]{2016ApJ...832...41M}
{Mordasini}, C., {van Boekel}, R., {Molli{\`e}re}, P., {Henning}, T., \&
  {Benneke}, B. 2016, \apj, 832, 41

\bibitem[{{Murgas} {et~al.}(2020){Murgas}, {Chen}, {Nortmann}, {Palle}, \&
  {Nowak}}]{2020A&A...641A.158M}
{Murgas}, F., {Chen}, G., {Nortmann}, L., {Palle}, E., \& {Nowak}, G. 2020,
  \aap, 641, A158

\bibitem[{{Nikolov} {et~al.}(2018){Nikolov}, {Sing}, {Fortney}, {Goyal},
  {Drummond}, {Evans}, {Gibson}, {De Mooij}, {Rustamkulov}, {Wakeford},
  {Smalley}, {Burgasser}, {Hellier}, {Helling}, {Mayne}, {Madhusudhan},
  {Kataria}, {Baines}, {Carter}, {Ballester}, {Barstow}, {McCleery}, \&
  {Spake}}]{2018Natur.557..526N}
{Nikolov}, N., {Sing}, D.~K., {Fortney}, J.~J., {et~al.} 2018, \nat, 557, 526

\bibitem[{{{\"O}berg} {et~al.}(2011){{\"O}berg}, {Murray-Clay}, \&
  {Bergin}}]{2011ApJ...743L..16O}
{{\"O}berg}, K.~I., {Murray-Clay}, R., \& {Bergin}, E.~A. 2011, \apjl, 743, L16

\bibitem[{{Pearson} {et~al.}(2019){Pearson}, {Griffith}, {Zellem}, {Koskinen},
  \& {Roudier}}]{2019AJ....157...21P}
{Pearson}, K.~A., {Griffith}, C.~A., {Zellem}, R.~T., {Koskinen}, T.~T., \&
  {Roudier}, G.~M. 2019, \aj, 157, 21

\bibitem[{{Rustamkulov} {et~al.}(2023){Rustamkulov}, {Sing}, {Mukherjee},
  {May}, {Kirk}, {Schlawin}, {Line}, {Piaulet}, {Carter}, {Batalha}, {Goyal},
  {L{\'o}pez-Morales}, {Lothringer}, {MacDonald}, {Moran}, {Stevenson},
  {Wakeford}, {Espinoza}, {Bean}, {Batalha}, {Benneke}, {Berta-Thompson},
  {Crossfield}, {Gao}, {Kreidberg}, {Powell}, {Cubillos}, {Gibson}, {Leconte},
  {Molaverdikhani}, {Nikolov}, {Parmentier}, {Roy}, {Taylor}, {Turner},
  {Wheatley}, {Aggarwal}, {Ahrer}, {Alam}, {Alderson}, {Allen}, {Banerjee},
  {Barat}, {Barrado}, {Barstow}, {Bell}, {Blecic}, {Brande}, {Casewell},
  {Changeat}, {Chubb}, {Crouzet}, {Daylan}, {Decin}, {D{\'e}sert},
  {Mikal-Evans}, {Feinstein}, {Flagg}, {Fortney}, {Harrington}, {Heng}, {Hong},
  {Hu}, {Iro}, {Kataria}, {Kempton}, {Krick}, {Lendl}, {Lillo-Box}, {Louca},
  {Lustig-Yaeger}, {Mancini}, {Mansfield}, {Mayne}, {Miguel}, {Morello},
  {Ohno}, {Palle}, {Petit dit de la Roche}, {Rackham}, {Radica},
  {Ramos-Rosado}, {Redfield}, {Rogers}, {Shkolnik}, {Southworth}, {Teske},
  {Tremblin}, {Tucker}, {Venot}, {Waalkes}, {Welbanks}, {Zhang}, \&
  {Zieba}}]{2023Natur.614..659R}
{Rustamkulov}, Z., {Sing}, D.~K., {Mukherjee}, S., {et~al.} 2023, \nat, 614,
  659

\bibitem[{{Seager} \& {Sasselov}(2000)}]{2000ApJ...537..916S}
{Seager}, S. \& {Sasselov}, D.~D. 2000, \apj, 537, 916

\bibitem[{{Sing} {et~al.}(2011){Sing}, {D{\'e}sert}, {Fortney}, {Lecavelier Des
  Etangs}, {Ballester}, {Cepa}, {Ehrenreich}, {L{\'o}pez-Morales}, {Pont},
  {Shabram}, \& {Vidal-Madjar}}]{2011A&A...527A..73S}
{Sing}, D.~K., {D{\'e}sert}, J.~M., {Fortney}, J.~J., {et~al.} 2011, \aap, 527,
  A73

\bibitem[{{Sing} {et~al.}(2016){Sing}, {Fortney}, {Nikolov}, {Wakeford},
  {Kataria}, {Evans}, {Aigrain}, {Ballester}, {Burrows}, {Deming},
  {D{\'e}sert}, {Gibson}, {Henry}, {Huitson}, {Knutson}, {Lecavelier Des
  Etangs}, {Pont}, {Showman}, {Vidal-Madjar}, {Williamson}, \&
  {Wilson}}]{2016Natur.529...59S}
{Sing}, D.~K., {Fortney}, J.~J., {Nikolov}, N., {et~al.} 2016, \nat, 529, 59

\bibitem[{{Skaf} {et~al.}(2020){Skaf}, {Bieger}, {Edwards}, {Changeat},
  {Morvan}, {Kiefer}, {Blain}, {Zingales}, {Poveda}, {Al-Refaie}, {Baeyens},
  {Gressier}, {Guilluy}, {Jaziri}, {Modirrousta-Galian}, {Mugnai}, {Pluriel},
  {Whiteford}, {Wright}, {Yip}, {Charnay}, {Leconte}, {Drossart}, {Tsiaras},
  {Venot}, {Waldmann}, \& {Beaulieu}}]{2020AJ....160..109S}
{Skaf}, N., {Bieger}, M.~F., {Edwards}, B., {et~al.} 2020, \aj, 160, 109

\bibitem[{{Sotzen} {et~al.}(2020){Sotzen}, {Stevenson}, {Sing}, {Kilpatrick},
  {Wakeford}, {Filippazzo}, {Lewis}, {H{\"o}rst}, {L{\'o}pez-Morales}, {Henry},
  {Buchhave}, {Ehrenreich}, {Fraine}, {Garc{\'\i}a Mu{\~n}oz}, {Jayaraman},
  {Lavvas}, {Lecavelier des Etangs}, {Marley}, {Nikolov}, {Rathcke}, \&
  {Sanz-Forcada}}]{2020AJ....159....5S}
{Sotzen}, K.~S., {Stevenson}, K.~B., {Sing}, D.~K., {et~al.} 2020, \aj, 159, 5

\bibitem[{{Southworth}(2011)}]{2011MNRAS.417.2166S}
{Southworth}, J. 2011, \mnras, 417, 2166

\bibitem[{{Spake} {et~al.}(2021){Spake}, {Sing}, {Wakeford}, {Nikolov},
  {Mikal-Evans}, {Deming}, {Barstow}, {Anderson}, {Carter}, {Gillon}, {Goyal},
  {Hebrard}, {Hellier}, {Kataria}, {Lam}, {Triaud}, \&
  {Wheatley}}]{2021MNRAS.500.4042S}
{Spake}, J.~J., {Sing}, D.~K., {Wakeford}, H.~R., {et~al.} 2021, \mnras, 500,
  4042

\bibitem[{{Sun} {et~al.}(2024){Sun}, {Xuesong Wang}, {Welbanks}, {Teske}, \&
  {Buchner}}]{2024arXiv240208292S}
{Sun}, Q., {Xuesong Wang}, S., {Welbanks}, L., {Teske}, J., \& {Buchner}, J.
  2024, arXiv e-prints, arXiv:2402.08292

\bibitem[{{Trotta}(2008)}]{2008ConPh..49...71T}
{Trotta}, R. 2008, Contemporary Physics, 49, 71

\bibitem[{{Visscher} {et~al.}(2010){Visscher}, {Lodders}, \&
  {Fegley}}]{2010ApJ...716.1060V}
{Visscher}, C., {Lodders}, K., \& {Fegley}, Bruce, J. 2010, \apj, 716, 1060

\bibitem[{{Welbanks} \& {Madhusudhan}(2019)}]{2019AJ....157..206W}
{Welbanks}, L. \& {Madhusudhan}, N. 2019, \aj, 157, 206

\bibitem[{{Welbanks} {et~al.}(2019){Welbanks}, {Madhusudhan}, {Allard},
  {Hubeny}, {Spiegelman}, \& {Leininger}}]{2019ApJ...887L..20W}
{Welbanks}, L., {Madhusudhan}, N., {Allard}, N.~F., {et~al.} 2019, \apjl, 887,
  L20

\bibitem[{{Wilson} {et~al.}(2021){Wilson}, {Gibson}, {Lothringer}, {Sing},
  {Mikal-Evans}, {de Mooij}, {Nikolov}, \& {Watson}}]{2021MNRAS.503.4787W}
{Wilson}, J., {Gibson}, N.~P., {Lothringer}, J.~D., {et~al.} 2021, \mnras, 503,
  4787

\bibitem[{{Zacharias} {et~al.}(2015){Zacharias}, {Finch}, {Subasavage},
  {Bredthauer}, {Crockett}, {Divittorio}, {Ferguson}, {Harris}, {Harris},
  {Henden}, {Kilian}, {Munn}, {Rafferty}, {Rhodes}, {Schultheiss}, {Tilleman},
  \& {Wieder}}]{2015AJ....150..101Z}
{Zacharias}, N., {Finch}, C., {Subasavage}, J., {et~al.} 2015, \aj, 150, 101

\end{thebibliography}
%



\end{document}